\documentclass[%
reprint,
amsmath,
amssymb,
aps,
prx,
superscriptaddress,
longbibliography,
nobalancelastpage,
]{revtex4-2}

\usepackage{graphicx}\usepackage{dcolumn}\usepackage{bm}\usepackage{hyperref}\usepackage[svgnames]{xcolor}

\usepackage{physics2}
\usephysicsmodule[tightbraces=true]{ab}
\usephysicsmodule{braket,doubleprod}

\makeatletter
\def\textemdash{\leavevmode\unskip\kern1pt---\kern1pt\ignorespaces}
\let\old@Section@Cmd=\section
\def\end@of@sec@tit{.\unskip---\kern1pt\ignorespaces}%
\def\mysection{%
  \def\reserved@stsec##1{\@startsection{section}{1}{\parindent}{\z@}{-0pt}{\normalfont\normalsize\itshape}*[##1]{##1\end@of@sec@tit}}%
  \@ifstar{%
    \reserved@stsec%
  }{%
    \reserved@stsec%
  }%
}
\let\section=\mysection
\makeatother

\usepackage[separate-uncertainty=false,per-mode=symbol,print-unity-mantissa = false]{siunitx}
\DeclareSIUnit\gauss{G}
\DeclareSIUnit\centimeter{cm}

\usepackage{braket}

\def\refapp#1{Appendix~\ref{#1}}%

\def\mqvgates{}

\hypersetup{
    pdftitle = {Recoil-free Quantum Gates with Optical Qubits},
    pdfauthor = {Zhao Zhang, Léo Van Damme, Marco Rossignolo, Lorenzo Festa, Max Melchner, Robin Eberhard, Dimitrios Tsevas, Kevin Mours, Eran Reches, Johannes Zeiher, Sebastian Blatt, Immanuel Bloch, Steffen Glaser, and Andrea Alberti},
	colorlinks=true,linkcolor=blue,citecolor=blue,filecolor=blue,urlcolor=blue
  }

\newcommand{\figref}[2]{\hyperref[#1]{\ref{#1}(#2)}}

\let\hbarorig=\hbar
\def\hbar{\hbarorig}

\def\ketq#1{\ket{#1}_\text{q}}
\def\ketm#1{\ket{#1}_\text{m}}

\def\recfree{rec.\hspace{1.3pt}free}

\def\LMU{Fakultät für Physik, Ludwig-Maximilians-Universität München, 80799 München, Germany}
\def\MCQST{Munich Center for Quantum Science and Technology, 80799 München, Germany}
\def\MPQ{Max-Planck-Institut für Quantenoptik, 85748 Garching, Germany}
\def\TUM{School of natural sciences, Technical University of Munich, 85747 Garching, Germany}
\def\QRUISE{Qruise GmbH, Saarbrücken, 66113, Germany}

\begin{document}

\title{Recoil-free Quantum Gates with Optical Qubits}

\author{Zhao Zhang}
\thanks{These authors contributed equally to this work.}
\affiliation{\MPQ}
\affiliation{\MCQST}

\author{L\'eo Van Damme}
\thanks{These authors contributed equally to this work.}
\affiliation{\TUM}
\affiliation{\MCQST}

\author{Marco Rossignolo}
\affiliation{\QRUISE}

\author{Lorenzo Festa}
\affiliation{\MPQ}
\affiliation{\MCQST}

\author{Max Melchner}
\affiliation{\MPQ}
\affiliation{\LMU}
\affiliation{\MCQST}

\author{Robin Eberhard}
\affiliation{\MPQ}
\affiliation{\LMU}
\affiliation{\MCQST}

\author{Dimitrios Tsevas}
\affiliation{\MPQ}
\affiliation{\LMU}
\affiliation{\MCQST}

\author{Kevin Mours}
\affiliation{\MPQ}
\affiliation{\LMU}
\affiliation{\MCQST}

\author{Eran Reches}
\affiliation{\MPQ}
\affiliation{\LMU}
\affiliation{\MCQST}

\author{Johannes Zeiher}
\affiliation{\MPQ}
\affiliation{\LMU}
\affiliation{\MCQST}

\author{Sebastian Blatt}
\affiliation{\MPQ}
\affiliation{\LMU}
\affiliation{\MCQST}

\author{Immanuel Bloch}
\affiliation{\MPQ}
\affiliation{\LMU}
\affiliation{\MCQST}

\author{Steffen J. Glaser}
\affiliation{\TUM}
\affiliation{\MCQST}

\author{Andrea Alberti}\email{andrea.alberti@mpq.mpg.de}
\affiliation{\MPQ}
\affiliation{\LMU}
\affiliation{\MCQST}

\date{\today}

\begin{abstract}
We propose a scheme to perform optical pulses that suppress the effect of photon recoil by three orders of magnitude compared to ordinary pulses in the Lamb-Dicke regime.
We derive analytical insight about the fundamental limits to the fidelity of optical qubits for trapped atoms and ions.
This paves the way towards applications in quantum computing for realizing $>1000$ of gates with an overall fidelity above $\SI{99}{\percent}$.
\end{abstract}

\keywords{Neutral-atom-based quantum computing}

\maketitle

\section{Introduction}

Ultranarrow optical transitions enable today's most accurate clocks due to their long coherence times and large energy level splitting compared to clocks based on atomic microwave transitions 
\cite{Ludlow:2015,Takamoto:2005,Bloom:2014,McGrew:2018,Norcia:2019a,Madjarov:2019,Brewer:2019,Young:2020,Bothwell:2022,Aeppli:2024,Finkelstein:2024}.
Can the same ultranarrow optical transitions be used to define a qubit for quantum computing applications which is competitive in terms of speed and fidelity with other atomic qubit implementations
\cite{Harty:2014,Srinivas:2021,Levine:2019,Evered:2023}
in trapped ions \cite{Leibfried:2003,Haffner:2008,Bruzewicz:2019,Monroe:2021}
and neutral atoms \cite{Saffman:2010,Saffman:2016,Adams:2020,Kaufman:2021,Wu:2021,%
Bluvstein:2022,Graham:2022,Bluvstein:2024}?
The answer is not obvious since optical qubits and optical clocks work in entirely different parameter regimes. 
The most striking example is the speed of operation.  The optical qubit has to be manipulated on a very short time scale \cite{Lis:2023}, whereas optical clock transitions are naturally probed over a long period of time for a higher resolution \cite{Ludlow:2015}.
This raises the challenge of how to operate the optical qubit at high Rabi frequencies while 
controlling the light shift induced by the driving laser (so-called probe shift \cite{Taichenachev:2006,Hu:2020}) and suppressing the effect of photon recoil in a regime where the motional sidebands are not inhibited.
The second fundamental difference in the requirements is the need for an optical qubit to operate a universal set of gates from arbitrary initial states \cite{Nielsen:2010}, as opposed to optical clocks for which spectroscopy is performed from a well-defined initial state \cite{Ludlow:2015}.
In fact, the optical qubit requires understanding and optimizing the quantum process of the gate, rather than only controlling a particular state evolution \cite{Lis:2023}.
Thus, the goal for an optical qubit is to implement arbitrary, fast quantum gates while controlling both recoil and probe shift.

The effect of photon recoil has long been studied \cite{Leibfried:2003}. There are two well-established methods to suppress it. One can work in the regime of low Rabi frequency (sideband resolved regime), and/or engineer deep traps (Lamb-Dicke regime), for which the photon recoil is to a great extent absorbed by the trapping potential instead of the atom (Mößbauer effect).
Both regimes are naturally approached with increasing trap frequency \cite{Bergquist:1987,Ido:2003}.
However, there are limits to the maximum trapping frequency for ions and especially for neutral atoms, where for the latter the trapping frequency is limited by the laser power and by photon scattering of the trapping light \cite{Saffman:2005}.
For neutral-atom quantum computing, efforts have been made to understand the role of photon recoil \cite{Robicheaux:2021} and to mitigate its effects on optical qubits \cite{Lis:2023}.
However, this problem has not been studied systematically nor have general solutions for arbitrary gates been found so far.
Other solutions have been proposed to suppress photon recoil by driving a three-photon transition. %
This, however, requires excellent phase coherence between the three laser beams and
precise control of their beam direction and polarization \cite{Hong:2005,Ryabtsev:2011,Barker:2016}.

\begin{figure}[b]
	\centering
	\includegraphics[scale=1]{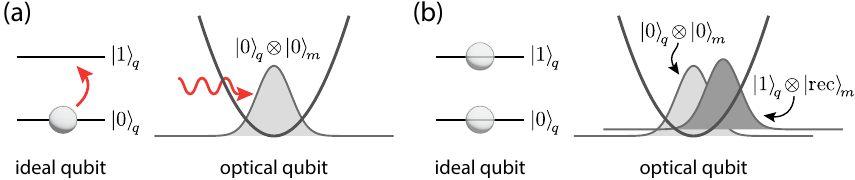}
	\caption{Ideal qubit vs.\ trapped optical qubit: (a) An ideal qubit is driven from $\ketq0$ to a superposition state, whereas the optical qubit absorbs the photon momentum when excited. (b) Instead of realizing the intended superposition of $\ketq0$ and $\ketq1$ states, the photon momentum causes a recoil from the initial motional state $\ketm{0}$ to $\ketm{\text{rec}}$ for the atom ending in state $\ketq1$, thus producing detrimental entanglement between internal and external degrees of freedom.}
	\label{fig:1}
\end{figure}

In this paper, we derive a new understanding of the physics underlying photon recoil, systematically quantify its impact on the gate operations on optical qubits and, based on the new insights, design a recoil-free pulse scheme for quantum gates. 
Leveraging the novel recoil-free pulse, we develop a composite pulse protocol to drive arbitrary gates on the optical qubit.
The new composite pulse scheme provides a unifying solution to all three relevant challenges: (1) we employ recoil-free pulses to suppress motional decoherence, (2) we fully parallelize slow operations on the optical qubit transition while performing local operations through fast $\sigma_z$ rotations, (3) we enhance the recoil-free pulses to what we later call Mikado pulses in order to be insensitive to probe shift, and correct systematic unitary errors by acting on the local operations.

\section{Recoil-free gates}
Optical qubits are fundamentally different from an ideal two-level system, as illustrated in Fig.~\ref{fig:1}:
The photon inducing the transition in an optical qubit carries a momentum $\hbar k$ that may not be negligible compared to the width of the momentum distribution of the trapped atom.
The effect of recoil manifests itself in two specific ways: entanglement of the qubit state with the motional state, leading to a direct reduction in the gate fidelity, and motional heating of the atom reducing the fidelity of subsequent gates indirectly.

We set the stage by introducing the Hamiltonian governing the interaction between an optical qubit and a resonant laser,
\begin{multline}
	\hat{H}[\varphi(t)]=\frac{\hbar\Omega}{2}(e^{i\varphi(t)+i\eta(\hat{a}^\dag+\hat{a})}\hat{\sigma}^++\text{H.c.})+\hbar\omega \hat{a}^\dag \hat{a} = \\
    \frac{\hbar\Omega}{2} [\hat{h}_0{+}\eta(a^\dag{+}\hat{a})\hat{h}_1{-}\frac{\eta^2}{2}(\hat{a}^\dag{+}\hat{a})^2\hat{h}_0]{+}\hbar\omega \hat{a}^\dag \hat{a} {+} \mathcal{O}(\eta^3),
\label{eqn:Hamiltonian_clock}
\end{multline}
which acts on the product space $\mathcal{Q}\otimes\mathcal{M}$ defined by the qubit ($\mathcal{Q}$) and motional states ($\mathcal{M}$).
In $\hat{H}$, we omit terms dealing with inhomogeneities of the probe shift, which are discussed later, and only focus on the photon-recoil effect, which is truly fundamental.
The control parameters in the Hamiltonian are the positive Rabi frequency $\Omega$, the phase $\varphi(t)$ of the laser field modulated in time $t$, and the trap frequency $\omega$.
The Hamiltonian also contains the Lamb-Dicke parameter $\eta = k x_0 = k\sqrt{\hbar/(2m\omega)}$ with $x_0$ being the zero-point width of the atom in the trap and $m$ its mass, the Pauli matrices $\hat{\sigma}$ with standard index notation, and the annihilation operator $\hat{a}$.
In Eq.~(\ref{eqn:Hamiltonian_clock}), a series expansion of $\hat{H}$ is provided to the order of $\eta^2$, where $\hat{h}_0=\hat{\sigma}_x\cos\varphi(t)+\hat{\sigma}_y\sin\varphi(t)$ and $\hat{h}_1=\hat{\sigma}_y\cos\varphi(t)-\hat{\sigma}_x\sin\varphi(t)$ are operators on $\mathcal{Q}$, whose relevance will become clear in the following.

\begin{figure}[t]
	\centering
	\includegraphics[width=\columnwidth]{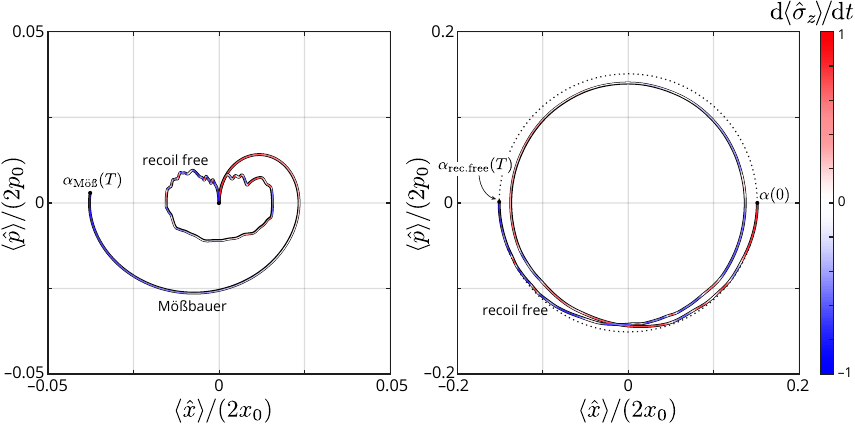}
	\caption{Phase space representation of the atom's motional state. The gate $\hat{R}_x({\pi/2})$ is applied to the initial qubit state $\sin(\pi/8)\ketq{0}+i\cos(\pi/8)\ketq{1}$. Left: Mößbauer pulse vs.\ recoil-free pulse for the initial motional Fock state $\ketm{0}$. Right: recoil-free pulse applied to a coherent state with $\alpha_\text{\recfree}(0) {=} 0.15$. The final state $\alpha_\text{\recfree}(T)$ is the same as if the atom had no pulse applied. The dotted circle with radius $|\alpha(0)|$ is provided for reference.}
	\label{fig:2}
\end{figure}
We obtain a basic understanding of the effect of photon-recoil by studying the motion of an atom in phase space.
To present our results, we will consider, without loss of generality, the specific case of a $\pi/2$ $\sigma_x$-rotation $\hat{R}_x(\pi/2)$, and first realize it with a constant phase $\varphi(t)=0$.
We call this a Mößbauer pulse because the transferred momentum in the Lamb-Dicke regime is suppressed to the order of $\hbar k\,(\Omega/\omega)/\sqrt{6}$ according to the Mößbauer effect (\refapp{appendix:moessbauer}).
This residual recoil effect can be completely eliminated by introducing recoil-free gates, which modulate the phase  $\varphi(t)$ to bring the motional state back to the origin in phase space: a formal definition of recoil-free pulses and how to compute them will be provided later.
In Fig.~\ref{fig:2}, we show a comparison of the Mößbauer and recoil-free gates for a particular initial qubit state.
To the first order of $\eta$, the recoil-free pulse not only suppresses the recoil for the particular state in the figure, but (i) for all initial qubit states, (ii) for motional Fock states, and (iii) even for coherent states of any amplitude $\alpha$.
This constitutes one of the main results of this paper.

To explain the statements (i)-(iii), we derive equations of motions for the $\hat{x}(t)$ and $\hat{p}(t)$ position and momentum operators in the Heisenberg picture.
These are the Newtonian equations of motion for a driven harmonic oscillator,
\begin{align}
\label{eqn:phase_space_eom_p}
\partial_t \hat{p}(t) &=-m\omega^2 \hat{x}(t)-\hbar k\,\partial_t \!\left(\!\frac{\hat{\sigma}_z}{2}\! \right),\\[-6pt]
\label{eqn:phase_space_eom_x}
\partial_t \hat{x}(t) &= \hat{p}(t)/m,\\
\label{eqn:phase_space_eom_qubit}
\partial_t\hat{\sigma}_z &= \Omega[\hat{\sigma}_y\cos\varphi(t)-\hat{\sigma}_x\sin\varphi(t)] + \mathcal{O}(\eta)
\end{align}
where Eqs.~(\ref{eqn:phase_space_eom_p}) and (\ref{eqn:phase_space_eom_x}) follow from an approximation of $\hat{H}[\varphi]$ in Eq.~(\ref{eqn:Hamiltonian_clock}) to the first order in $\eta$, and Eq.~(\ref{eqn:phase_space_eom_qubit}) describes the qubit dynamics to zero order in $\eta$ (\refapp{appendix:phase_space}).
From these equations, an intuitive, semiclassical picture emerges, where the trajectory in phase space is determined by the interplay between the harmonic oscillator force and the recoil force applied to the atom, which is given by the excitation rate $\partial_t \braket{\hat{\sigma}_z}/2$.
In the following, we sketch a proof of three statements (\refapp{appendix:symmetries_qubit} for details):
(i) The temporal phase profile 
$\varphi(t)$ is designed to suppress recoil for the three initial states: $\ket{0}$, $(\ket{0}+\ket{1})/\sqrt{2}$ and $(\ket{0}+i\ket{1})/\sqrt{2}$.  
Owing to the linearity of the Newtonian equations above and the time-reversal symmetry of $\hat{H} = - \hat{\tau}^{-1}\hat{H} \hat{\tau}$, with $\hat{\tau}$ the time-reversal operator, if $\ket{0}$ evolves under the gate in a recoil-free way, so does $\hat{\tau} \ket{0}=\ket{1}$ too.
Combining these results, we conclude that any qubit superposition state is also recoil free, $\braket{\hat{x}(T)}=\braket{\hat{p}(T)}=0$.
(ii) Because the expectation value of $\hat{x}$ and $\hat{p}$ vanishes for any Fock state, and the zeroth-order qubit dynamics is independent of motion, the driving force is independent of the initial Fock state.
(iii) Given the linear structure of Eqs.~(\ref{eqn:phase_space_eom_p}) and (\ref{eqn:phase_space_eom_x}), the generic trajectory for an initial coherent state $\alpha_\text{\recfree}(t)$ can be decomposed into two solutions: the homogeneous solution $\alpha_\text{hom}(t)$ when no pulse is applied and the particular solution with the recoil-free pulse starting and ending at $\alpha=0$.
Hence, $\alpha_\text{\recfree}(T) = \alpha_\text{hom}(T)$.

While the semi-classical picture explains the suppression of motional excitations, it does not capture the entanglement between the qubit and the motional states, which ultimately limits the gate fidelity.
To advance beyond the phase-space representation, we perform quantum process tomography \cite{Nielsen:2010} of the unitary $\hat{U}$ generated by $\hat{H}$ and use the results to evaluate the fidelity against a target unitary $\hat{U}_\text{tar}$ acting on $\mathcal{Q}$.
By taking the partial trace over $\mathcal{M}$, we obtain the operator-sum representation of $\hat{U}$ in $\mathcal{Q}$:
\begin{equation}
	\mathcal{E}_{\hat{U}\hspace{-1.5pt},\hspace{1pt}\rho_m}\!(\hat{\rho}_q)=\text{Tr}_m[\hat{U}(\hat{\rho}_q\otimes\hat{\rho}_m) U^\dag]=\sum_{k=0}^{3} \chi_k \hat{E}_k\hat{\rho}_q\hat{E}_k^\dag,
	\label{eqn: process_tomography}
\end{equation}
where $\hat{E}_i$ are the Kraus operators and $\chi_i$ are the probabilities of the different quantum channels, where the dominant channel with index $0$ obeys $1{-}\chi_0\ll 1$.
In addition, $\rho_m$ denotes the initial motional state, which we choose to be a thermal state with ground state probability $p_0$.
By applying the formalism in Ref.~\cite{Pedersen:2007}, we compute the fidelity of the process $\mathcal{E}$ averaged over the initial qubit state as $\langle\mathcal{F}\rangle= \big[1+2\sum_k\chi_k|\text{Tr}(\hat{U}_{\text{tar}}^\dag \hat{E}_k)/2|^2\big]/3$.
Importantly, we find that the infidelity $J=1-\langle\mathcal{F}\rangle$ can be bounded from above by $J_\text{ent} {+} J_\text{uni}$, with the two contributions,
\begin{align}
	\label{eq:J_ent}
	J_{\text{ent}}&=\frac{2}{3}(1-\chi_0),\\
	\label{eq:J_uni}
	J_{\text{uni}}&=\frac{2}{3}(1-|\text{Tr}(\hat{U}_{\text{tar}}^\dag \hat{E}_0)/2|^2),
\end{align}
being interpreted as the infidelity by entanglement and by systematic deviations from $\hat{U}_\text{tar}$.
Note that to derive the bound, we only consider $\hat{E}_0$.
While $J$ captures the process infidelity, it does not account for motional heating, which, if not suppressed, leads to a lower $p_0$ (i.e., higher temperature) for the subsequent gates and, thus, to a higher infidelity.
Hence, we introduce an additional cost function $J_\text{mot}$ defined as the change of $\braket{\hat{a}^\dagger \hat{a}}$ in absolute value, averaged over the four initial qubit states used in the process tomography (\refapp{numerical_optimization_rfpulse}).

\begin{figure}[t]
            	\centering
	\includegraphics[width=\columnwidth]{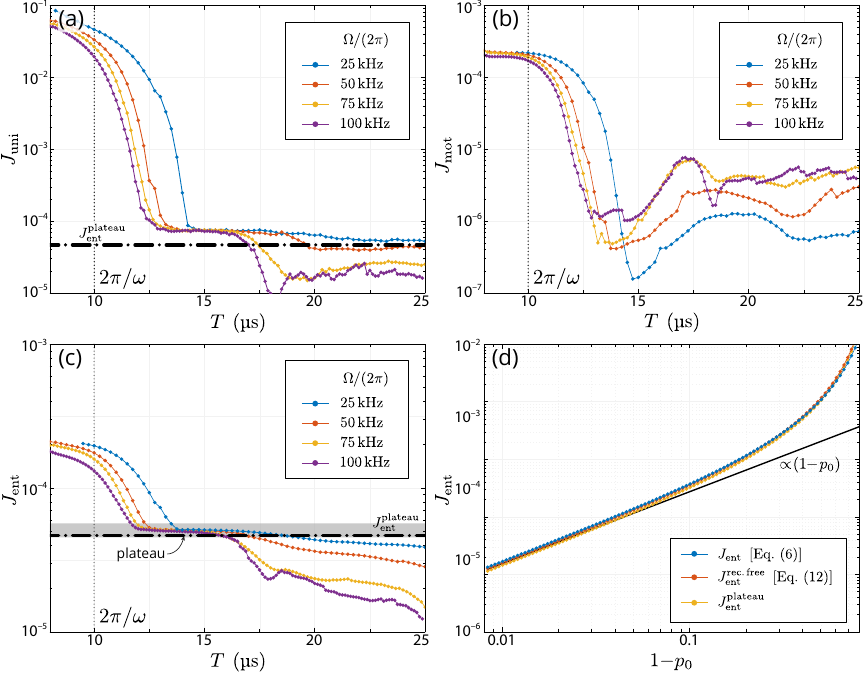}
	\caption{Quantum process tomography of recoil-free pulses as a function of pulse duration $T$, for $\omega=2\pi\times \SI{100}{\kilo\hertz}$, $\eta=0.22$, $p_0=0.95$. (a) $J_\text{uni}$, (b) $J_\text{mot}$, (c) $J_\text{ent}$. (d)
	$J_\text{ent}$ as a function of the ground state probability $p_0$ for $T=\SI{15}{\micro\second}$.}
	\label{fig:3}
\end{figure}

To obtain the recoil-free pulses, we modulate $\varphi(t)$ while minimizing the weighted sum of the cost terms $J_\text{uni}$, $J_\text{ent}$ and $J_\text{mot}$ for $\hat{U}_\text{tar}=\hat{R}_x(\pi/2)$.
We present in Figs.~\figref{fig:3}{a-c} the three terms as a function of the pulse duration $T$ for different values of $\Omega$.
The first relevant observation is that the three cost terms drop for durations $>2\pi/\omega$, indicating that the quantum speed limit is determined by the trap frequency.
Second, we find that $J_\text{mot}$ is increasingly suppressed as we decrease the Rabi frequency, as is expected in the sideband resolved regime.
However, the same behavior is not observed for the other cost functions $J_\text{ent}$ and $J_\text{uni}$, which plateau once the recoil (i.e., $J_\text{mot}$) is suppressed.

To explain the results above, we expand the unitary $\hat{U}$ acting on $\mathcal{Q}\otimes \mathcal{M}$ to the second order in $\eta$: $\hat{U}(T)=\hat{U}_0(T)[1+\eta \hat{V}_1(T)+\eta^2 \hat{V}_2(T)+\mathcal{O}(\eta^3) ]$, with the following definitions:
\begin{align}
\hat{U}_0(t)=&\hat{U}_q(t)  e^{-i a^\dagger \hspace{-1pt} a\,\omega t},\\
\hat{V}_1(t)=& \hat{V}_\text{rec}^{(1)}(t)\, a^{\dagger}{-}\text{H.c.},\\
\hat{V}_2(t)=&\big( \hat{V}_\text{rec}^{(2)}(t)\,\hat{a}^{\dag 2}-\text{H.c.}\big)-i\,\hat{V}_\text{ent}^{(2)}(t)\,\hat{a}^\dag\hat{a}.
\end{align}
Here, $\hat{U}_q(t)$ is the unitary generated by the Hamiltonian $\frac{\hbar\Omega}{2}\left(1-\frac{\eta^2}{2}\right)\hat{h}_0(t)$, which yields the zeroth-order evolution of the ideal qubit under a renormalized Rabi frequency.
In addition, $\hat{V}_\text{rec}^{(1)}(t)=-i\frac{\Omega}{2}\int_0^t\hat{U}_q(\tau)\hat{h}_1(\tau)\hat{U}_q^\dag(\tau)e^{i\omega \tau}\mathrm{d}\tau$ and $\hat{V}_\text{rec}^{(2)}(t)=-i\frac{\Omega}{4}\int_0^t\hat{U}_q(\tau)\hat{h}_0(\tau)\hat{U}_q^\dag(\tau)e^{2i\omega \tau}\mathrm{d}\tau$ are responsible for the first- and second-order motional heating with an exchange of one and two motional quanta, respectively.
They can be understood as the Fourier transform of $\hat{h}_1$ and $\hat{h}_0$ in the rotating frame, evaluated at frequency $\omega$ and $2\omega$.
In contrast, the operator $\hat{V}_\text{ent}^{(2)}(t)$ does not change the motional state and is responsible for entanglement, which will be discussed later.
To obtain $J_\text{mot}=0+\mathcal{O}(\eta^5)$ (i.e., suppress recoil to the order of $\eta^4$) requires the conditions $\hat{V}_\text{rec}^{(1)}(T)=\hat{V}_\text{rec}^{(2)}(T)=0$.
To fulfill these conditions, $T$ must be larger than one trap period: $T>2\pi/\omega$, since phase-modulated pulses with duration $T$ can only control the spectrum of the operator at frequencies larger than $2\pi/T$.
Hence, this establishes the quantum speed limit $T_\text{QSL}$ of recoil-free pulses.

When the recoil-free condition $\hat{V}_\text{rec}^{(1)}(T)=\hat{V}_\text{rec}^{(2)}(T)=0$ is held, $J_\text{ent}$ is only determined by $\hat{V}_\text{ent}^{(2)}(t)=\frac{\Omega}{2}\int_0^t \hat{U}_q(\tau)\hat{h}_0(\tau)\hat{U}_q^\dag(\tau)\mathrm{d}\tau$.
Using process tomography, we evaluate $J_\text{ent}$ to the order of $\eta^4$ for recoil-free pulses (\refapp{appendix:thermal_quantum_channel}):
\begin{equation}
	\label{eq:Jent_rec_free}
    J_\text{ent}^\text{\recfree}\approx \frac{2}{3}\frac{1-p_0}{p_0}\eta^4 \big|\vec{B}(\hat{V}_\text{ent}^{(2)}(T))\big|^2 +\mathcal{O}(\eta^6),
\end{equation}
where $\vec{B}$ defines a mapping from a Hermitian operator $\hat{A}$ to a vector: $\vec{B}(\hat{A}) = \{\text{Tr}(\hat{A}\hat{\sigma}_x),\text{Tr}(\hat{A}\hat{\sigma}_y),\text{Tr}(\hat{A}\hat{\sigma}_z)\}/2$.
To evaluate this expression, we make the approximation $\hat{U}_q(t) \approx \hat{R}_x(\Omega t)$, which holds because $\hat{V}_\text{ent}^{(2)}$ is the dc component of $\hat{h}_0$ in the rotating frame and, thus, is marginally affected by the high-frequency components of the phase modulation in a recoil-free pulse, whose spectrum comprises frequencies above $2\pi/T$.
With this approximation, we obtain $J_\text{ent}^\text{plateau} = \frac{\pi^2}{24}\frac{1-p_0}{p_0}\eta^4$, which explains the plateau in Fig.~\figref{fig:3}{c}, independent of the Rabi frequency $\Omega$.
For sufficiently long durations, however, the phase modulation allows controlling lower and lower frequency components and, thus, reducing $J_\text{ent}^\text{\recfree}$ below the plateau value.
We emphasize that such a suppression of $\hat{V}_\text{ent}^{(2)}$ is only possible because of the modulation of $\varphi$ and could not be achieved by modulating $\Omega$ alone.
In fact, if the phase is kept fixed, $\hat{h}_0$ would commute with $\hat{U}_q$, implying that $V_{\text{ent}}^{(2)}$ would be proportional to the area of the pulse, which needs to be greater than $\pi/2$ and thus cannot vanish.
Importantly, this fact shows that even when operating deep in the sideband resolved regime, constant-phase pulses cannot reduce $J_\text{ent}$ below $J_\text{ent}^\text{plateau}$.
The second important result we obtain from Eq.~(\ref{eq:Jent_rec_free}) is the linear scaling with $1{-}p_0$, provided a sufficiently small temperature, as shown in Fig.~\figref{fig:3}{d}.

\section{Composite pulse protocol for universal one-qubit gates}
\begin{figure}[t]

	\centering
	\includegraphics[width=\columnwidth]{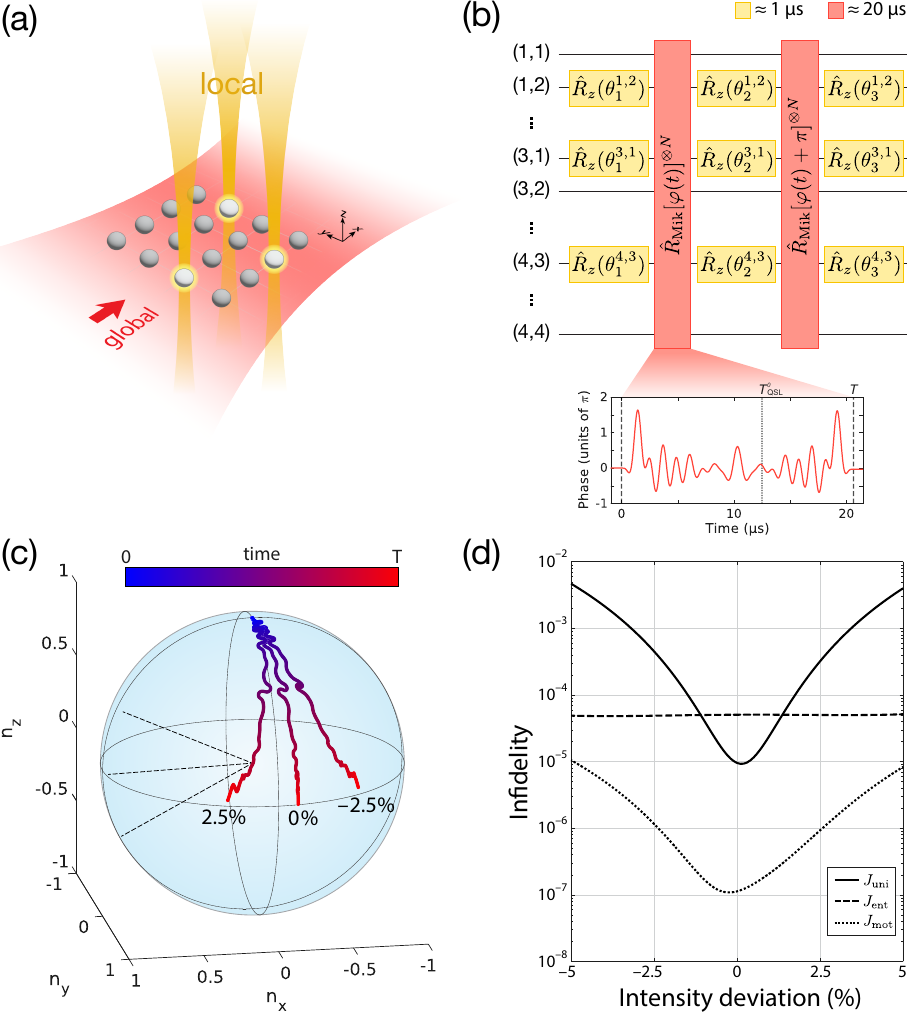}
	\caption{(a) Laser configuration with global recoil-free pulses and local $\sigma_z$-rotations by local light shifts. (b) Composite pulse scheme based on global Mikado pulses and site-dependent $R_z(\theta)$ rotations. Inset: phase modulation, with duration $\SI{65}{\percent}$ longer than zeroth-order quantum speed limit $T_\text{QSL}^0 = \pi/(2\Omega)$. (c) Bloch vector trajectory of Mikado pulse under intensity deviation $\delta I/I = \{-0.025, 0, 0.025\}$.
	The dashed lines represent the rotation axis of the effective rotation by the Mikado pulse for the three intensities, respectively. (d) Three cost functions vs.\ intensity deviation for the Mikado pulse. In the figure, the Mikado pulse is optimized for ${}^{88}$Sr atoms; see \refapp{appendix:Sr_setup}.}
	\label{fig:4}
\end{figure}
We show that recoil-free pulses enable fast, arbitrary quantum gates for quantum computing with optical qubits.
The purpose is to implement an arbitrary unitary gate $\hat{U}_g \in \text{SU(2)}$.
Naively, this could be implemented with a Euler decomposition \cite{Wang:2016}, $\hat{R}_x(\theta_3)\hat{R}_y(\theta_2)\hat{R}_x(\theta_1)$, where each rotation is recoil free.
This can be realized with parallel addressing of the atoms, relying on scalable opto-electronic solutions \cite{Zhang:2024,Graham:2023,Menssen:2023,Christen:2022}, which is within reach, but not demonstrated with atoms yet.
Instead, we propose an alternative scheme that avoids addressing the atoms sequentially and, thus, avoids long execution time pulses because of the relatively small Rabi frequency, typical of ultranarrow optical transitions.
The proposed scheme only uses two recoil-free gates per circuit step, based on a global recoil-free pulse, together with fast, local $\hat{\sigma}_z$-rotations $\hat{R}_z(\theta)$ implemented by light shift on the individual atoms; see Fig.~\figref{fig:4}{a}.
Inspired by the Euler decomposition, the composite scheme consists in applying the following operations:
\begin{equation}
\begin{aligned}
    \hat{U}_g=&\hat{R}_z(\theta_3)\hat{R}_\text{Mik}[\varphi(t)+\pi] \hat{R}_z(\theta_2)\hat{R}_\text{Mik}[\varphi(t)]\hat{R}_z(\theta_1).
\end{aligned}
\label{eqn: Pulse_composite}
\end{equation}
as schematically illustrated in Fig.~\figref{fig:4}{b}.
Here, $\hat{R}_\text{Mik}[\varphi(t)]$ is a class of pulses transforming the north pole of the Bloch sphere to its equator, which we call Mikado after the popular children's game:
\begin{equation}
    \hat{R}_\text{Mik}[\varphi(t)]=\hat{R}_z(\alpha)\hat{R}_x(\pi/2)\hat{R}_z(\beta).
    \label{eqn: Pulse_Mikado_decomposition}
\end{equation}

\begin{figure*}[t]
	\centering
	\includegraphics[scale=1]{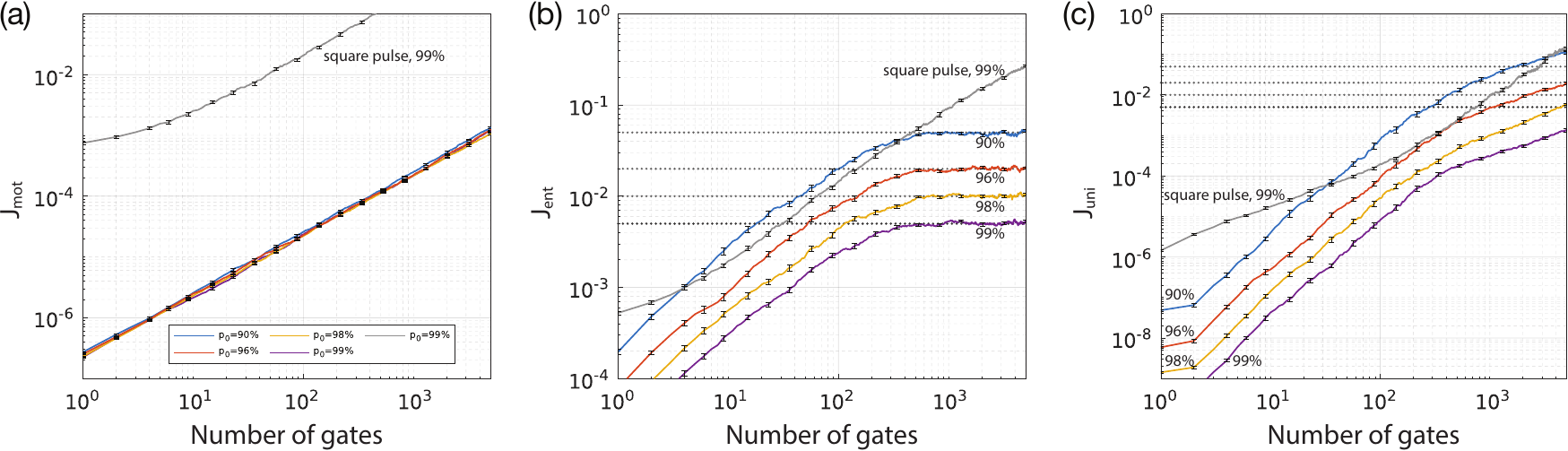}
	\caption{Randomized benchmarking of a sequences of SU(2) gates drawn from the Haar measure, realized with the composite pulse of Fig.~\figref{fig:4}{b}, for different ground state probabilities $p_0$ (see legend), whereas the gray curves refer to constant-phase Mößbauer pulses for $p_0=\SI{99}{\percent}$.
    The saturation value of $J_\text{ent}$ (see text) is shown in both (b) and (c) as a point of comparison (dotted lines).
	Representative 1-$\sigma$ error bars are shown.
	\label{fig:5}}
\end{figure*}

Mikado pulses are generated like recoil-free pulses by modulating $\varphi(t)$ to minimize the three cost functions $J_\text{uni}^\text{Mik}$, $J_\text{ent}$ and $J_\text{mot}$, where $J_\text{uni}^\text{Mik}$ is adapted from $J_\text{uni}$ (Appendix~\ref{appendix:Optimization_Mikado}).
Thereby, Mikado pulses inherit the same recoil-free characteristic.
However, they allow for two loose parameters, $\alpha$ and $\beta$, in their optimization, which serves as additional degrees of freedom.
Compared to a recoil-free fixed rotation $\hat{R}_x(\pi/2)$, Mikado pulses are robust against spatial inhomogeneities of $\Omega$ and detuning $\hbar\delta\,\hat{\sigma}_z/2$, which in realistic scenarios need to be added to $\hat{H}$ in Eq.~(\ref{eqn:Hamiltonian_clock}).
An example of inhomogeneous detuning is the probe shift from the intensity inhomogeneity of the global pulse.
The idea behind Mikado pulses is to map site-dependent deviations of $\Omega$ and $\delta$ in $\hat{H}$ onto the deviations $\delta\alpha$ and $\delta\beta$ of the two loose parameters.
The robustness of Mikado pulses is exemplified by the three different Bloch-sphere trajectories shown in Fig.~\figref{fig:4}{c}, corresponding to three different laser intensities.
Figure \figref{fig:4}{d} shows in addition that the recoil-free property (i.e., vanishing $J_\text{ent}$ and $J_\text{mot}$) is preserved despite the parameter inhomogeneity.
Hence, knowing the parameters for each site allows correcting the deviations $\alpha+\delta \alpha $ and $\beta+\delta \beta$ in the Mikado gates by suitable $\sigma_z$-rotation angles.

The composite pulse scheme allows implementing any gate $\hat{U}_g$ preserving comparable values of $J_\text{ent}$ and $J_\text{mot}$ as for Mikado pulses [Fig.~\figref{fig:4}{d}], while suppressing $J_\text{uni}$ to values below $10^{-6}$ owing to the local $\hat{R}_z$ rotations; see \refapp{appendix:infidelity_colormap}.
Importantly, the high fidelity of single gates carries over to long sequences, as shown by the randomized benchmarking in Fig.~\ref{fig:5}.
In particular, $J_\text{mot}$ in Fig.~\ref{fig:5} \hyperref[fig:5]{(a)} shows an improvement by at least three orders of magnitude in  heating suppression compared to Mößbauer pulses.
Such a strong suppression is key to achieve excellent $J_\text{ent}$ and $J_\text{uni}$ for a scalable number of gates, as shown in Fig.~\ref{fig:5}\hyperref[fig:5]{(b)} and Fig.~\ref{fig:5}\hyperref[fig:5]{(c)}.
Remarkably, $J_\text{ent}$ exhibits a saturation effect to $(1{-}p_0)/2$, derived in \refapp{appendix:randomized_benchmarking}.
This asymptotic behavior underlines the importance of preparing sufficiently cold atoms to carry out high-fidelity gates on optical qubits.
Remarkably, even in the conservative scenario of ground-state  population $p_0=\SI{90}{\percent}$, the crossover point defined by $J_\text{uni}>J_\text{ent}$ is reached for a number of gates above 1000.

\section{Conclusions}

In this paper, we have shown that our recoil-free pulses, enhanced by the Mikado scheme, provide an affirmative answer to the opening question: they enable applications of optical qubits for quantum computing and entanglement-enhanced quantum metrology with state-of-the-art fidelities.

\begin{acknowledgments}
We acknowledge support from QRUISE in simulating and optimizing the recoil-free pulses.
We thank Rainer Blatt and Xie-Hang Yu for insightful discussions, and Tobias Olsacher for careful reading of the manuscript.
We acknowledge funding from Munich Quantum Valley project TAQC, the BMBF project MUNIQC-Atoms, and the Munich Center for Quantum Science and Technology.
M.\hspace{1pt}R.\ acknowledges funding from PASQUANS2.1.
\end{acknowledgments}

\bibliographystyle{apsrev4-2}

\cleardoublepage
\appendix
\makeatletter
\def\appendixname{APPENDIX}%
\def\section#1{\old@Section@Cmd{\uppercase{#1}}}

\section{Analyzing Atom Motion in Phase Space}
\label{appendix:phase_space}
In this section, we derive the Heisenberg equations of motion for a trapped atom subject to an optical drive.
We show that these take the form of semi-classical equations of motion of a driven harmonic oscillator, providing us with a tool to intuitively understand the recoil-free dynamics of quantum gates.
Moreover, this also offers a straightforward explanation for extending the recoil-free properties of optimized pulses from Fock states to coherent states. 

The Hamiltonian of Eq.~(\ref{eqn:Hamiltonian_clock}) up to first order in $\eta$ is:
\begin{equation}
\begin{aligned}
    &\hat{H}/\hbar=\frac{\Omega}{2}[\hat{\sigma}_x\cos\varphi(t)+\hat{\sigma}_y\sin\varphi(t)]+\omega \hat{a}^\dag \hat{a} \\
    &+\frac{\eta\Omega}{2}(\hat{a}^\dag+\hat{a})[\hat{\sigma}_y\cos\varphi(t)-\hat{\sigma}_x\sin\varphi(t)]+\mathcal{O}(\eta^2)
\end{aligned}
\end{equation}
The Heisenberg equations of motion for the operators $\hat{\sigma}_z$ and $\hat{a}$ under this Hamiltonian are:
\begin{equation}
\begin{aligned}
    \partial_t\hat{\sigma}_z &= \Omega[\hat{\sigma}_y\cos\varphi(t)-\hat{\sigma}_x\sin\varphi(t)]+\mathcal{O}(\eta),\\
    \partial_t \hat{a} &= -i\omega \hat{a}-i\frac{\eta\Omega}{2}[\hat{\sigma}_y\cos\varphi(t)-\hat{\sigma}_x\sin\varphi(t)]+\mathcal{O}(\eta^2).
\end{aligned}
\end{equation}
By substituting the equation of motion for 
$\hat{\sigma}_z$ into that for $\hat{a}$, we derive:
\begin{equation}
    \partial_t \hat{a}= -i\omega \hat{a} -i\eta \;\partial_t\!\left( \frac{\hat{\sigma}_z}{2}\right)
\end{equation}
From this equation, we directly obtain the equation of motion for $\hat{x}$ and $\hat{p}$:
\begin{equation}
\begin{aligned}
\label{eq:x_p_eq_motion}
\partial_t \hat{p} =-m&\omega^2 \hat{x}-\hbar k\partial_t (\frac{\hat{\sigma}_z}{2})\\
\partial_t \hat{x} &= \hat{p}/m
\end{aligned}
\end{equation}
This equation of motion can be understood as the atom moving in a harmonic potential while subject to the force,
\begin{equation}
    \label{eq:driving_force}
    \hat{f} = -\hbar k\partial_t \left(\frac{\hat{\sigma}_z}{2}\right),
\end{equation}
originating from the momentum transfer (i.e., the recoil).
The force is determined by the zeroth-order qubit dynamics, more specifically, by the transition rate between the two qubit states.

\section{Mößbauer pulses}
\label{appendix:moessbauer}

We refer to pulses with constant phase $\varphi$ as Mößbauer pulses for the reasons provided in the text. 
These pulses provide a useful reference to be compared with the recoil-free gate.
In this section, we compute the phase space dynamics associated with the atomic motional state during a Mößbauer pulse.
The computation makes use of the equations of motions derived in Appendix~\ref{appendix:phase_space}, which give an approximation of the dynamics in the Lamb-Dicke regime to the first order of $\eta$.

We parametrize the Bloch vector of the initial qubit state with spherical coordinates $\{\theta,\phi\}$:
\begin{eqnarray}
\langle \hat{\sigma}_x(0)\rangle &=&\cos\theta,\\
\langle \hat{\sigma}_y(0)\rangle &=&\sin\theta\sin\phi,\\
\langle \hat{\sigma}_z(0)\rangle &=&\sin\theta\cos\phi,
\end{eqnarray}
with the $x$-axis defining the zenith direction.
For the evaluation of the driving force in Eq.~(\ref{eq:driving_force}), we compute $\langle \hat{\sigma}_z(t)\rangle$ under the application of a Mößbauer gate ($\varphi(t)=0$):
\begin{equation}
    \langle \hat{\sigma}_z(t) \rangle= \sin\theta\cos (\phi-\Omega t).
\end{equation}
Using this expression in  Eq.~(\ref{eq:x_p_eq_motion}), we derive 
\begin{equation}
    \label{eq:newton}
    \Ddot{x}(t)+\omega^2 x(t) = f(t)/m,
\end{equation}
where $x(t)$ is the expectation value $\braket{\hat{x}}$ and $f(t) =\braket{\hat{f}}$ is the expectation value of the driving force.

With the initial condition $x(0)=0$ and $\Dot{x}(0)=0$, the solution of  Newton's equation (\ref{eq:newton}) is:
\begin{eqnarray}    
    x(t) &=& -x_0\eta\sin\theta\bigg[\frac{\omega\Omega}{\omega^2-\Omega^2}[\sin(\Omega t-\phi) \nonumber \\
    && +\sin\phi\cos(\omega t)-\frac{\Omega}{\omega}\cos\phi\sin(\omega t)\bigg],\\
    p(t) &=& -p_0\eta\sin\theta\frac{\omega\Omega}{\omega^2-\Omega^2}\bigg[\frac{\Omega}{\omega}\cos(\Omega t-\phi) \nonumber \\
    && -\sin\phi\sin(\omega t)-\frac{\Omega}{\omega}\cos\phi\cos(\omega t)\bigg].
\end{eqnarray}
where $p(t)=m\Dot{x}(t)$.

We specialize the solution to the case considered in the text of a $\pi/2$ pulse defined by $T=\pi/(2\Omega)$.
The motional state at the end of the pulse is:
\begin{eqnarray}  
    \frac{x(T)}{x_0}&=&-\eta\sin\theta \frac{1}{\xi^2-1}\bigg[\xi\cos\phi \nonumber\\
    &&+\xi\sin\phi\cos(\frac{\pi\xi}{2})-\cos\phi\sin(\frac{\pi\xi}{2})\bigg]\\
     \frac{p(T)}{p_0}&=&-\eta\sin\theta \frac{1}{\xi^2-1}\bigg[\sin\phi \nonumber \\
    &&-\xi\sin\phi\sin(\frac{\pi\xi}{2})-\xi^2\cos\phi\cos(\frac{\pi\xi}{2})\bigg],
\end{eqnarray}  
where $\xi=\omega/\Omega$.

Using these solutions of the motion equation, we compute the amplitude of the coherent state average over the four qubit states considered in the quantum process tomography:
\begin{eqnarray}
   \braket{|\alpha (T)|^2} &=& \frac{\braket{x(T)}^2}{(2x_0)^2}+\frac{\braket{p(T)}^2}{(2p_0)^2} \\
    &=&\frac{3}{16} \frac{\xi ^2-2 \xi  \sin \left(\frac{\pi  \xi }{2}\right)+1}{ \left(\xi ^2-1\right)^2}\,\eta^2.
\end{eqnarray}
Neglecting in this expression the small oscillating term, the order of magnitude of the transferred momentum can be approximated as $2p_0  \sqrt{\braket{|\alpha (T)|^2}}=\hbar k\,\xi/\sqrt{6}$, which is given in the text.

\section{Symmetries of Phase Space Dynamics}
\label{appendix:symmetries_qubit}

We use the semiclassical equations of motion in Eqs.~(\ref{eqn:phase_space_eom_p}) and (\ref{eqn:phase_space_eom_x}) (see Appendices~\ref{appendix:phase_space} and \ref{appendix:moessbauer} for the derivation) to prove the claims (i), (ii), and (iii) in the text, namely:
the recoil-free pulse condition obtained for selected qubit states and the Fock state $\ketm{0}$ extends directly to all qubit states and motional Fock and coherent states.
These claims are based on the same assumptions used to derive the semiclassical equations of motion in Eq.~(\ref{eq:newton}), notably retaining in the equations of motions only first-order terms in $\eta$.
We also use the same notation introduced in Appendix~\ref{appendix:moessbauer} to denote the expectation value of $\hat{x}$ and $\hat{f}$ with $x(t)$ and $f(t)$.

Firstly, we prove claim (i): if the recoil-free condition is held for the initial qubit state $\ketq{0}$, $(\ketq{0}+\ketq{1})/\sqrt{2}$ and $(\ketq{0}+i\ketq{1})/\sqrt{2}$, then it is held for all initial qubit states.
The proof relies on the linearity of the Newtonian equation Eq.~(\ref{eq:newton}) and of the Schrödinger equation.
In fact, by the linearity of Eq.~(\ref{eq:newton}), if $x_i(t)$ is the solution for the force $f_i(t)$, then $x(t)=\sum_i c_i x_i(t)$ is the solution for the sum of the forces, $f(t)=\sum_i c_if_i(t)$.
Therefore, if under the force $f_i(t)$, the system is recoil-free at the end of the pulse (i.e., $x(T)=0$ and $\Dot{x}(T)=0$), then under an arbitrary linear superposition of force the system is also recoil-free.
We also notice that the force $f(t)$ is solely determined by the zero-order dynamics in Eq.~(\ref{eqn:phase_space_eom_qubit}), representing the Schrödinger equation for the qubit.
Thus, if the initial density matrix $\rho_i$ results in the force $f_i(t)$, then the superposition of density matrices $\sum_i c_i \rho_i(t)$ yields the sum of the forces, $f(t)=\sum_i c_if_i(t)$.
Therefore, if $\rho_i$ produces a force $f_i$ with recoil-free dynamics, then the superposition of density matrices also produces a force with recoil-free dynamics, according to the argument provided above.
Based on this result, we conclude that if the system is recoil-free for the complete basis set of qubit density matrices, that is, for the four density matrices defined by the pure states $\ketq{0}$, $\ketq{1}$ $(\ketq{0}+\ketq{1})/\sqrt{2}$ and $(\ketq{0}+i\ketq{1})/\sqrt{2}$, then it is recoil-free for all initial qubit states.

Moreover, we can relax the previous requirements by imposing the recoil-free condition on only three of the four qubit states defined above:
in fact, if the pulse is recoil free for the initial state $\ketq{0}$, we can show that it is as well for the initial state $\ketq{1}$.
This follows from the symmetry of the Hamiltonian, $\hat{\tau}^{-1}\hat{H}\hat{\tau}=-\hat{H}$, under the time-reversal operation $\hat{\tau} = i\sigma_y K$, where $K$ is the complex conjugation operator in the basis where $\hat{\sigma}_z$ is diagonal.
In turn, the antisymmetry of $\hat{H}$ implies that the unitary evolution operator $\hat{U}$ is invariant under $\tau$, 
$\hat{\tau}^{-1} \hat{U}\hat{\tau}=\hat{U}$.
Based on this, if a pair of initial qubit states are time-reversal conjugated, $\hat{\tau}\ketq{0}=\ketq{1}$, then the two states are also time-reversal conjugated after time evolution, $\hat{\tau}\hat{U}\ketq{0}=\hat{U}\ketq{1}$.
Hence, the forces $f_0(t)$ and $f_1(t)$ associated with the two initial states $\ketq{0}$ and $\ketq{1}$ are directly related, $f_1(t)=-f_0(t)$, because of the definition of the force in Eq.~(\ref{eq:driving_force}) and $\hat{\tau}^{-1}\hat{\sigma}_z\hat{\tau}=-\hat{\sigma}_z$.
Since $f_0(t)$ is recoil-free, $f_1(t)$ is as well.

We prove claim (ii): if the recoil-free condition is fullfilled for the ground motional state $\ketm{0}$, it is also fullfilled for arbitrary motional Fock states $\ketm{n}$. 
This directly follows from the fact that all Fock states give the same initial condition in the semi-classical picture: $x(0)=\bra{n}\hat{x}\ket{n}=0$, $\Dot{x}(0)=\bra{n}\hat{p}\ket{n}/m=0$. 
Therefore, according to the semiclassical equation of motion, the trajectory in phase space goes back to the origin for all initial Fock states because they share the same initial conditions.

Finally, we prove claim (iii): if a pulse satisfies the recoil-free conditions for the ground motional state $\ketm{0}$, its effect on a coherent motional state $\ketm{\alpha}$ is equivalent to a free propagation for a pulse duration $T$. 
The phase space trajectory for an atom prepared in $\ketm{0}$ represents a particular recoil-free solution of the inhomogeneous  Eq.~(\ref{eq:newton}), satisfying $x_\text{ground}(0)=x_\text{ground}(T)=0$, $\Dot{x}_\text{ground}(0)=\Dot{x}_\text{ground}(T)=0$. 
The phase space trajectory for an atom prepared in $\ketm{\alpha}$ is also determined by Eq.~(\ref{eq:newton}), with the same driving force $f(t)$ but with different initial conditions $x_\text{coh}(0)=2x_0 \text{Re}(\alpha)$, $\Dot{x}_\text{coh}(0)=2p_0 \text{Im}(\alpha)/m$.
The solution of the motion equations of the coherent state can be expressed as the sum of the particular solution $x_\text{ground}$ and a solution $x_\text{free}(t)$ of the homogeneous equation defined by the non-driven (i.e., $f(t)=0$) harmonic oscillator: $x_\text{coh}(t)=x_\text{ground}(t)+x_\text{free}(t)$.
Thus, at the end of the pulse, the motional state of the atom is $x_\text{coh}(T)=x_\text{ground}(T)+x_\text{free}(T)=x_\text{free}(T)$, since the ground state dynamics is assumed to be recoil free, $x_\text{ground}(T)=0$.
This represents the same point in phase space that is reached by the coherent state after evolving for a time $T$ in the harmonic trap, free of any drive.

The arguments provided above for (ii) and (iii) prove that for a recoil-free pulse, the final motional state in phase space (i.e, the centroid of the motional state distribution) is not affected by the pulse.
This leaves open the possibility that the motional state distribution is distorted by the pulse, while its centroid follows the prediction by the semiclassical equation Eq.~(\ref{eq:newton}).
However, the terms that distort the distribution are of the order of $\eta^2$ and higher.
These terms are true quantum mechanical effects, which are not captured by the trajectory in phase space and relate to higher-order recoil effects and qubit-motion entanglement, both discussed in Appendices~\ref{appendix:unitary_expansion_1} and \ref{appendix:unitary_expansion_2}.

\section{Numerical Optimization for a recoil-free pulse}
\label{numerical_optimization_rfpulse}
We present the method used for optimizing recoil-free pulses in Fig.~\ref{fig:2} and Fig.~\ref{fig:3}.
The cost function is defined as the contribution of three different terms:
\begin{equation}
    J=w_\text{ent}J_\text{ent}+w_\text{uni}J_\text{uni}+w_\text{mot}J_\text{mot},
\label{eq:weighted_sumed_J}
\end{equation}
where $w_\text{ent}$, $w_\text{uni}$ and $w_\text{mot}$ are the weight of corresponding cost function terms, whereas $J_\text{ent}$ is the infidelity originating from qubit-motion entanglement and  $J_\text{uni}$ is the systematic error from the target unitary, with their definition being provided in Eq.~(\ref{eq:J_ent}) and Eq.~(\ref{eq:J_uni}).
The third contribution in Eq.~(\ref{eq:weighted_sumed_J}) is $J_\text{mot}$, representing the change of $a^\dag a$ in absolute value, averaged over the four initial qubit states used in the process tomography:
\begin{equation}
    J_\text{mot}=\frac{1}{4}\sum_{k=1}^4\sum_{n=0}^\infty p_n\big|\text{Tr}[(\hat{U}\hat{\rho}_n^k\hat{U}^\dag-\hat{\rho}_n^k)\hat{a}^\dag \hat{a}]\big|,
\end{equation}
where $\rho_n^k=\ket{\psi_n^k}\bra{\psi_n^k}$ defines the density matrix of the initial state $\ket{\psi_n^k}=\ketq{\psi_{k}}\otimes \ketm{n}$ in the space $\mathcal{Q}\otimes \mathcal{M}$.
Here, $\ketq{\psi_{k}}$ represents one of the four qubit states, $|0\rangle$, $|1\rangle$, $(|0\rangle+|1\rangle)/\sqrt{2}$, $(|0\rangle+i|1\rangle)/\sqrt{2}$, whereas $\ketm{n}$ denotes the $n$-th Fock state.
The coefficients $p_n=p_0(1-p_0)^n$ represent the occupation probability of the $n^\text{th}$ motional Fock state under the Boltzmann distribution.
We use the weights $w_\text{ent}=100$, $w_\text{uni}=1$ and $w_\text{mot}=100$ for Fig.~\ref{fig:2}, and $w_\text{ent}=100$, $w_\text{uni}=1$ and $w_\text{mot}=10$ for Fig.~\ref{fig:3}.

The phase-modulation pulse is constructed as the product of a Fourier series and a regularization mask: $\varphi(t)=u(t)\mu(t)$, where 
\begin{equation}
u(t)=\sum_{n=1}^{N_c}a_n\cos\left(n\pi\tfrac{t}{T}\right)+b_n\sin\left(n\pi\tfrac{t}{T}\right),
\end{equation}
while $N_c$ is the cut-off number of Fourier coefficients, which is chosen between 40 and 60. The regulation mask avoids abrupt discontinuities in $\varphi(t)$ and is defined by:
\begin{equation}
\begin{aligned}
\mu(t)=
&\begin{aligned}
\begin{cases}
\displaystyle
\frac{1-\cos\left(\tfrac{10\pi t}{T}\right)}{2}& \text{ if } t < T/10,\\
1 & \text{ if } T/10\leq t\leq 9T/10,\\
\displaystyle
\frac{1+\cos\left(\tfrac{10\pi t}{T}\right)}{2}& \text{ if } t > 9T/10.
\end{cases}
\end{aligned}\\
\end{aligned}
\end{equation} 

All the Fourier coefficients ${a_n, b_n}$ are numerically optimized to minimize the cost function in Eq.~(\ref{eq:weighted_sumed_J}) and, thus, to obtain the recoil-free pulse.

\section{Perturbative Expansion of Unitary under Recoil}
\label{appendix:unitary_expansion_1}
In this section, we give a general perturbation expansion up to the second order in $\eta$ of the unitary $\hat{U}$ generated by the Hamiltonian in Eq.~(\ref{eqn:Hamiltonian_clock}).
This perturbation expansion provides the basis for the analysis provided in Fig.~\ref{fig:3}.
The expansion of the Hamiltonian up to the second order in $\eta$ is
\begin{equation}
    \hat{H}(t)=\hat{H}_0(t)+\eta \hat{H}_1(t)+\eta^2 \hat{H}_2(t)+\mathcal{O}(\eta^3),
\label{eqn:Hamiltonian_eta_square}
\end{equation}
where
\begin{eqnarray}
    \hat{H}_0(t) &=& \frac{\hbar\Omega}{2}(1-\frac{\eta^2}{2})\hat{h}_0(t)+ \hbar\omega \hat{a}^\dag\hat{a},
    \label{eqn:Hamiltonian_expansion_0}\\
    \hat{H}_1(t) &=& \frac{\hbar\Omega}{2}(\hat{a}^\dag+\hat{a})\hat{h}_1(t),
    \label{eqn:Hamiltonian_expansion_1}
    \\
    \hat{H}_2(t) &=& -\frac{1}{2}\frac{\hbar\Omega}{2} (\hat{a}^{\dag2}+\hat{a}^2)\hat{h}_0(t) -\frac{\hbar\Omega}{2} \hat{a}^\dag\hat{a}\hat{h}_0(t),
\label{eqn:Hamiltonian_expansion_2}
\end{eqnarray}
and
\begin{equation}
\begin{aligned}
    \hat{h}_0(t) &=\hat{\sigma}_x\cos\varphi(t)+\hat{\sigma}_y\sin\varphi(t),\\
    \hat{h}_1(t) &=\hat{\sigma}_y\cos\varphi(t)-\hat{\sigma}_x\sin\varphi(t).
\end{aligned}
\label{eqn:Hamiltonian_qubit}
\end{equation}

To understand the effect of recoil, we need to calculate the series expansion of $\hat{U}$ generated by this Hamiltonian.
For simplicity, we firstly discuss the first-order perturbation of the Hamiltonian in Eq.~(\ref{eqn:Hamiltonian_eta_square}), rewriting it in the form $\hat{H}(t)=\hat{H}_0(t)+\eta \hat{H}'(t)$ with $\hat{H}'(t)=\hat{H}_1(t)+\eta \hat{H}_2(t)$.
The expansion of the unitary to the first order in $\eta$ is:
\begin{multline}    
    \hat{U}(T) = \mathcal{T}\left[\exp{\left(-i\int_0^T \hat{H}(t)dt\right) }\right]\\
         = \hat{U}_0(T)[\hat{I}+\eta \hat{V}'(T)]+\mathcal{O}(\eta^2).
\label{eqn:unitary_expansion1}
\end{multline}
where we factorized the zeroth-order unitary evolution $\hat{U}_0(T)$ and its first-order correction, described by $\hat{V}'(T)$.
From the Schrödinger equation, $i\partial_t \hat{U}(t)=\hat{H}(t)\hat{U}(t)$, we get the propagation of each order of $\hat{U}(T)$.

The zeroth order term $\hat{U}_0(t)$  satisfies:
\begin{equation}
    i\partial_t \hat{U}_0(t) =\hat{H}_0(t)\hat{U}_0(t),
\end{equation}
whose solution is:
\begin{equation}
\label{eq:U_0_solution}
\hat{U}_0(T)=\hat{U}_q(T)\otimes \exp(-i\omega T \hat{a}^\dag\hat{a}),
\end{equation}
\begin{equation}
    \hat{U}_q(T) = \mathcal{T}\left[\exp{\left(-i\frac{\Omega}{2}(1-\frac{\eta^2}{2})\int_0^T \hat{h}_0(t)dt\right) }\right],
\end{equation}
is the propagation of an ideal qubit under the pulse with dressed Rabi frequency $\tilde\Omega = \Omega(1-\eta^2/2)$ \cite{Leibfried:2003}.
From the expression of $\hat{U}_0(T)$, we recognize that at the zeroth order in $\eta$, the qubit and motion are decoupled and both of them propagate as if there were no coupling between them.
For the convenience in the later computation of the higher-order dynamics, we introduce the operators:
\begin{equation}
\begin{aligned}
    \hat{h}_0^I(t)=\hat{U}_q^\dag (t)\hat{h}_0(t)\hat{U}_q(t),\\
    \hat{h}_1^I(t)=\hat{U}_q^\dag (t)\hat{h}_1(t)\hat{U}_q(t),
\end{aligned}
\end{equation}
which represent $\hat{h}_0(t)$ and $\hat{h}_1(t)$ in the interaction picture defined by $\hat{U}_q(t)$.

The propagation of $\hat{V}'(t)$ is defined by:
\begin{equation}
\label{eq:diff_V_prime}
     i\partial_t \hat{V}'(t)=\hat{U}_0^\dag(t)\hat{H}'(t)\hat{U}_0(t).
\end{equation}
By integrating Eq.~(\ref{eq:diff_V_prime}) and using the definitions in Eqs.~(\ref{eqn:Hamiltonian_expansion_1}) and (\ref{eqn:Hamiltonian_expansion_2}), we obtain a solution in the form:
\begin{multline}
    \hat{V}'(T)=[\hat{a}^\dag \hat{V}_\text{rec}^{(1)}(T) - \text{H.c.}]\\
    +\eta[\hat{a}^{\dag2}\hat{V}_\text{rec}^{(2)}(T)- \text{H.c.}]-i\eta\hat{a}^\dag \hat{a}\hat{V}_\text{ent}^{(2)}(T),
\label{eqn:V_formal}
\end{multline}
where the operators $\hat{V}_\text{rec}^{(1)}$, $\hat{V}_\text{rec}^{(2)}$, and $\hat{V}_\text{ent}^{(2)}(T)$ are defined below.
We note that both operators $\hat{V}_\text{rec}^{(1)}$ and $\hat{V}_\text{rec}^{(2)}$ are responsible for changes of motional states, and we refer to them as recoil terms, whereas the operator $\hat{V}_\text{ent}^{(2)}(T)$ maintains the same motional state and is responsible for pure entanglement of the qubit with the motion.

The expression of the first order recoil operator is:
\begin{equation}
    \hat{V}_\text{rec}^{(1)}(T)=-i\frac{\Omega}{2}\int_0^T\mathrm{d}t\, e^{i\omega t}\hat{h}_1^I(t),
\end{equation}
which can be understood as a Fourier transform of $\hat{h}_1$ in a rotation frame at frequency $\omega$.
The effect of the second-order recoil term $\hat{V}_\text{rec}^{(2)}$ is described by the operator:
\begin{equation}
    \hat{V}_\text{rec}^{(2)}(T)=-i\frac{\Omega}{4}\int_0^T \mathrm{d}t\, e^{2i\omega t} \hat{h}_0^I(t),
\end{equation}
which can be understood as the Fourier transform of the operator $\hat{h}_0$ in the rotating frame at frequency $2\omega$.
The effect of the qubit-motion entanglement due to recoil comes from the operator:
\begin{equation}
    \hat{V}_\text{ent}^{(2)}(T)=\frac{\Omega}{2}\int_0^T \mathrm{d}t\,  \hat{h}_0^I(t),
\end{equation}
which is the dc component of $\hat{h}_0$ in the rotating frame.
These results justify the series expansion provided in the text.

\section{Rigorous treatment of second-order terms}
\label{appendix:unitary_expansion_2}
We note that the unitary computed in Eq.~(\ref{eqn:unitary_expansion1}) is strictly valid only to the first order in $\eta$, whereas the expression of $\hat{V}'(t)$ introduces terms, $\hat{V}_\text{ent}^{(2)}(T)$ and $\hat{V}_\text{ent}^{(2)}(T)$, which are of second order in $\eta$.
In the previous section, the justification for keeping these higher-order terms in the expression of $\hat{U}(T)$ is based on physical intuition. Below, we provide a rigorous derivation of $\hat{U}(T)$ to the second order in $\eta$.

We extend the procedure used to derive Eq.~(\ref{eqn:unitary_expansion1}) to the second order in $\eta$, and obtain the expansion:
\begin{equation}
\begin{aligned}
    \hat{U}(T) &= \mathcal{T}\left[\exp{\left(-i\int_0^T \hat{H}(t)dt\right) }\right]\\
         &= \hat{U}_0(T)[\hat{I}+\eta \hat{V}_1(T)+\eta^2 \hat{V}_2(T)]+\mathcal{O}(\eta^3).
\end{aligned}
\label{eqn:unitary_expansion}
\end{equation}
The zeroth order dynamic $\hat{U}_0(t)$ is same as provided in the previous section in Eq.~(\ref{eq:U_0_solution}).
The first-order term $\hat{V}_1(t)$ obeys:
\begin{equation}
     i\partial_t \hat{V}_1(t)=\hat{U}_0^\dag(t)\hat{H}_1(t)\hat{U}_0(t) =\frac{\Omega}{2}\left[e^{i\omega t}\hat{a}^\dag \hat{h}_1^I(t)+\text{H.c.}\right],
     \label{eq:V_1_diff_eq}
\end{equation}
with the solution being:
\begin{equation}
    \hat{V}_1(T)=\hat{a}^\dag \hat{V}_\text{rec}^{(1)}(T) - \text{H.c.},
\label{eqn:V_1_formal}
\end{equation}
which describe the first-order recoil effect appearing in Eq.~(\ref{eqn:V_formal}).
As expected, both zeroth- and first-order terms, $\hat{U}_0(T)$ and $\hat{V}_1(T)$, obtained here coincide with those derived from the calculation in the previous section, which is rigorous to the first order in $\eta$.

The second-order term $\hat{V}_2(t)$ in the unitary follows the equation:
\begin{multline}
    i\partial_t \hat{V}_2(t)=\hat{U}_0^\dag(t)\hat{H}_2(t)\hat{U}_0(t)+\hat{U}_0^\dag(t)\hat{H}_1(t)\hat{U}_0(t)\hat{V}_1(t)
\end{multline}
Combining this expression with Eq.~(\ref{eq:V_1_diff_eq}), we obtain:
\begin{multline}    
    i\partial_t \hat{V}_2(t) = \hat{U}_0^\dag(t)\hat{H}_2(t)\hat{U}_0(t)\\+\frac{i}{2}\partial_t [\hat{V}_1(t)^2]+\frac{i}{2}[\partial_t \hat{V}_1(t),\hat{V}_1(t)],
\end{multline}
with the solution being:
\begin{multline}
    \hat{V}_2(T) = -i\int_0^T dt \hat{U}_0^\dag(t)\hat{H}_2(t)\hat{U}_0(t)\\
    + \frac{1}{2}\hat{V}_1(T)^2+\int_0^T dt [\partial_t \hat{V}_1(t),\hat{V}_1(t)]
\label{eqn:V_2_formal}
\end{multline}
The first line in Eq.~(\ref{eqn:V_2_formal}) includes the second-order recoil term and the atom-motion entanglement term in Eq.~(\ref{eqn:V_formal}).
The second line in Eq.~(\ref{eqn:V_2_formal}) contains terms to the second order in $\eta$ that are not contained in the previous section.
However, we show below that these terms do not change the physical picture presented in the text.

The first term, $\hat{V}_1(T)^2/2$, simply vanishes when the recoil-free condition, $\hat{V}_1(T)=0$, is fulfilled to the first order in $\eta$.
The second term in Eq.~(\ref{eqn:V_2_formal}) is:
\begin{multline}
    \int_0^T \mathrm{d}t\, [\partial_t\hat{V}_1(t),\hat{V}_1(t)] =\\ -i\delta \hat{V}^{(2)}
     -i \hat{a}^\dag \hat{a} \hat{V}_\text{ent}^{(2)'}
     +(\hat{a}^{\dag 2}\hat{V}_\text{rec}^{(2)'}(T)-\text{H.c.}),
\end{multline}
where we recognize three contributions associated with the following operators: $\delta \hat{V}$ is a motion-independent correction to the unitary, $\hat{V}_\text{ent}^{(2)'}$ is a purely motion-qubit entanglement term, and $\hat{V}_\text{rec}^{(2)'}(T)$ is a recoil term.

The motion-independent correction contribution is:
\begin{multline}
    \label{eq:extra_corr}
    \delta \hat{V}^{(2)'}(T) = i\frac{\Omega^2}{4}\int_0^T \mathrm{d}t_2\int_0^{t_2}\mathrm{d}t_1 \\
    \times (e^{-i\omega(t_2-t_1)} \hat{h}_1^I(t_2)\hat{h}_1^I(t_1)-\text{H.c.}).
\end{multline}
The purely motion-qubit entanglement term is:
\begin{multline}
    \label{eq:extra_ent}
    \hat{V}_\text{ent}^{(2)'}(T) = -i\frac{\Omega^2}{2}\int_0^T \mathrm{d}t_2\int_0^{t_2}\mathrm{d}t_1 \\\cos({\omega(t_1-t_2)})
     [\hat{h}_1^I(t_1),\hat{h}_1^I(t_2)].
\end{multline}
The recoil term is:
\begin{multline}
    \label{eq:extra_recoil}
    \hat{V}_\text{rec}^{(2)'}(T) = \frac{\Omega^2}{4}\int_0^T \mathrm{d}t_2\int_0^{t_2}\mathrm{d}t_1 \\ e^{-i\omega(t_1+t_2)}[\hat{h}_1^I(t_1), \hat{h}_0^I(t_2)].
\end{multline}

Importantly, all three terms in Eqs.~(\ref{eq:extra_corr}), (\ref{eq:extra_ent}) and (\ref{eq:extra_recoil}) are the Fourier transforms at frequency $\omega$ of the two-point correlators obtained from  $\hat{h}^I_0(t)$ and $\hat{h}^I_1(t)$.
Following the same argument presented in the text, all these terms can  be suppressed for gate durations $T$ above the quantum speed limit $T_\text{QSL}=2\pi/\omega$.

\section{Thermal Quantum Operation}
\label{appendix:thermal_quantum_channel}

When the recoil-free condition is held: $\hat{V}_\text{rec}^{(1)}(T)=\hat{V}_\text{rec}^{(2)}(T)=0$, the propagator $\hat{U}(T)$ to the order of $\eta^2$ can be simplified as:
\begin{equation}
\begin{aligned}
    \label{eq:unitary_thermal_channel}
    \hat{U}(T)&\approx \hat{U}_0(T)[1-i\eta^2\hat{a}^\dag \hat{a}\hat{V}_\text{ent}^{(2)}(T)] +\mathcal{O}(\eta^3)\\
    &\approx \hat{U}_0(T)\,\hat{U}_\text{ent}+\mathcal{O}(\eta^3).
\end{aligned}
\end{equation}
The term $\hat{U}_\text{ent}$ is a motion-dependent unitary rotation on the qubit and generate entanglement between qubit and motion:
\begin{equation}
    \hat{U}_\text{ent}=\exp[-i\eta^2\hat{a}^\dag \hat{a}\hat{V}_\text{ent}^{(2)}(T)]
\label{eqn:entanglement_U}
\end{equation}

To understand the entanglement generated by $\hat{U}_\text{ent}$, without loss of generality, we consider a motion-dependent unitary acting on the qubit of the following form:
\begin{equation}
    \label{eq:thermal_unitary}
    \hat{U}_\text{th}= \exp\left(i\frac{\theta}{2}\hat{a}^\dag \hat{a}\hat{\sigma}_\mathbf{n}\right) = \sum_{n=0} \hat{V}^n \otimes \ketbra{n}{n},
\end{equation}
where $\hat{V}$ is a unitary on the qubit and is parametrized as $\hat{V}=\exp{(i\theta\hat{\sigma}_{\mathbf{n}}/2)}$, with the Hermitian operator $\sigma_{\mathbf{n}}=n_x\sigma_x+n_y\sigma_y+n_z\sigma_z$ and $|\mathbf{n}|=n_x^2+n_y^2+n_z^2=1$.
The operator $\hat{V}$ produces a rotation of the qubit on the Bloch sphere by an angle $\theta$ around the axis $\mathbf{n}$.
When applying this unitary on a trapped atom with thermal motional state: $\hat{\rho}_q\otimes\hat{\rho}_m$, where $\hat{\rho}_m=\sum_n p_0\delta p ^n|n\rangle\langle n|$ represents  the thermal motional state with ground state probability $p_0=1-\delta p$.
We note that $\delta p$ is the probability to occupy a higher motional state and is typically a small number.
From $\hat{U}_\text{th}$, we can derive the quantum process by tracing out the motional states:
\begin{equation}
\begin{aligned}
    \mathcal{E}_\text{th}(\hat{\rho}_q) &= \text{Tr}_m(\hat{U}_\text{th}\,\hat{\rho}_q\otimes\hat{\rho}_m\, \hat{U}_\text{th}^\dag) \\
    &=\sum_{n=0} p_0\delta p ^n \hat{V}^n \rho_q \hat{V}^{\dag n}.
\end{aligned}
\label{eqn: thermal_channel}
\end{equation}
Note that $\hat{V}^n=\cos(n\theta/2)\hat{\sigma}_0+i\sin(n\theta/2)\hat{\sigma}_\mathbf{n}$ includes only two orthogonal terms: the qubit identity operator $\hat{\sigma}_0$ and $\hat{\sigma}_\mathbf{n}$.
Therefore, quantum process in Eq.~(\ref{eqn: thermal_channel}) can be represented as
\begin{equation}
    \mathcal{E}_\text{th}(\hat{\rho}_q)=\sum_{i,j=1}^{2}\chi_{ij}\hat{E}_i \hat{\rho}_q \hat{E}_j^\dag,
\end{equation}
where we recognize two Kraus operators $\hat{E}_0=\hat{\sigma}_0$ and $\hat{E}_1=\hat{\sigma}_\mathbf{n}$, with $\chi$ being a $2\times 2$ matrix defined by:
\begin{multline}
    \chi=\frac{p_0}{2} \sum_n
    \left(\begin{array}{cc}
        1+\cos(n\theta) & -i\sin(n\theta)   \\
        i\sin(n\theta) & 1-\cos(n\theta)   
    \end{array}\right) \delta p^n \\
=\frac{1}{2}+\frac{p_0}{2}
    \left(\begin{array}{cc}
        \text{Re}(\frac{1}{1-\delta p e^{i\theta}}) & -i\text{Im}(\frac{1}{1-\delta p e^{i\theta}})   \\
        i\text{Im}(\frac{1}{1-\delta p e^{i\theta}}) & -\text{Re}(\frac{1}{1-\delta p e^{i\theta}}) 
\end{array}\right).
\label{eqn: thermal_channel_term}
\end{multline}
The diagonalization of the matrix $\chi$ gives the two eigenvalues
\begin{equation}
	\chi_{\pm} = \frac{1}{2}\left(1 \pm p_0\left|\frac{1}{1-\delta p \,e^{i\theta}}\right|\right) 
\end{equation}
with $(\chi_+ + \chi_-)=1$ and $(1-\chi_+)\ll 1$ for $\delta p\ll 1$. Following the definition of $J_\text{ent}$ in Eq.~(\ref{eq:J_ent}), we obtain:
\begin{align}
    J_\text{ent} &= \frac{1}{3}\left(1-p_0 \left|\frac{1}{1-\delta p \, e^{i\theta}}\right|\right) \\
    \label{eq:approx1}
    & \approx \frac{1}{6}\frac{\delta p}{p_0}\theta^2 +\mathcal{O}(\delta p^2\theta^3)\\
    \label{eq:approx2}
    & \approx \frac{2}{3} \sin^2\left(\frac{\theta}{2}\right) \delta p+\mathcal{O}\left( \delta p ^2\right)\\
    &\approx \frac{1}{6}(\delta p )\theta^2+\mathcal{O}\left( \delta p ^2\right)+\mathcal{O}\left(\delta p^2 \theta^3\right).
\end{align}
where the approximations in Eqs.~(\ref{eq:approx1}) and (\ref{eq:approx2}) find applications in later expressions. 

To obtain an explicit expression for $J_\text{ent}$, we evaluate $\theta=-2\eta^2|\vec{B}(\hat{V}_\text{ent}^{(2)}(T))|$, where $\vec{B}$ defines the mapping from a Hermitian operator $\hat{A}$ to a vector:
$\vec{B}(\hat{A}) = \{\text{Tr}(\hat{A}\hat{\sigma}_x),\text{Tr}(\hat{A}\hat{\sigma}_y),\text{Tr}(\hat{A}\hat{\sigma}_z)\}/2$.
Therefore, we have obtained the important result that the entanglement infidelity of a recoil-free pulse depends solely on $p_0$ and $\hat{V}_\text{ent}^{(2)}$ through the expression:
\begin{equation}
\begin{aligned}
    J_\text{ent}^\text{\recfree}{=} &\frac{1}{3}\!\left(\!1{-}p_0\left|\frac{1}{1-\delta p e^{2i\eta^2|\vec{B}(\hat{V}_\text{ent}^{(2)}(T))|}}\right|\right) + \mathcal{O}(\delta p^2\eta^6) \\
    = &\frac{2}{3}\frac{\delta p}{p_0} \left|\vec{B}(\hat{V}_\text{ent}^{(2)}(T))\right|^2 \!\eta^4 +\mathcal{O}(\delta p^2\eta^6)
    \label{eq:J_ent_recoil_free}
\end{aligned}
\end{equation}
where the second equation is obtained from a series expansion neglecting terms of order of  $\delta p^2\eta^6$.

To develop an intuition about $J_\text{ent}$ in Eq.~(\ref{eq:J_ent_recoil_free}), we consider the simple example of a constant-phase pulse (Mößbauer pulse) for a total duration $T$.
In this case, the zeroth-order ideal qubit dynamics is described by a rotation on the Bloch sphere, $\hat{U}_q(t) = \hat{R}_x(\Omega t)$, and the corresponding entanglement operator is:
\begin{equation}
    \hat{V}_\text{ent}^{(2)}(t) = \frac{\Omega}{2}\int_0^t \!\mathrm{d}\tau\, e^{i\Omega \hat{\sigma}_x \tau/2}\sigma_x e^{-i\Omega \hat{\sigma}_x \tau /2} =\hat{\sigma}_x \frac{\Omega t}{2},
\end{equation}
with the vector length $|\vec{B}(\hat{V}_\text{ent}^{(2)}(T))|=\Omega T/2$. Therefore, the contribution to the entanglement infidelity from $\hat{V}_\text{ent}^{(2)}(t)$ is
\begin{equation}
    J_\text{ent}=\frac{1}{6}\frac{\delta p}{p_0}\eta^4(\Omega T)^2 + J_\text{ent,rec}.
\label{eqn:J_ent_constant}
\end{equation}
where $J_\text{ent,rec}$ stands for the contribution to $J_\text{ent}$ from $\hat{V}_\text{rec}^{(1)}$ and $\hat{V}_\text{rec}^{(2)}$. 
In fact, it should be noted that a Mößbauer pulse is in general not recoil free, meaning that that both motional heating operators $\hat{V}_\text{rec}^{(1)}$ and $\hat{V}_\text{rec}^{(2)}$ are nonzero and, thus, contribute to $J_\text{ent}$.
For the $\pi/2$-pulse scenario considered in the text, we obtain:
\begin{equation}
	J_\text{ent} = \frac{\pi^2}{24}\frac{\delta p}{p_0}\eta^4.
\end{equation}
Importantly, this expression is an approximation of the plateau value in Fig.~\figref{fig:3}{c}.

\section{Optimization of Mikado Pulses}
\label{appendix:Optimization_Mikado}
We discuss the optimization of the Mikado pulses, which are robust against deviations of the Hamiltonian parameters.
Such pulses have been studied for nuclear magnetic resonance (NMR) applications (composite pulses of type B3 \cite{Levitt:1986}) to transform one particular initial state to a statistical mixture of Bloch vectors whose azimuthal angle is not fixed, but rather depends on the local imperfection.
Note that in contrast to NMR applications, controlling the azimuthal angle is crucial to realize quantum gates.
Such a control of the azimuthal angle is possible with the sequence introduced in Eq.~(\ref{eqn: Pulse_composite}).

The error in the Mikado pulse is defined by the distance between the Mikado unitary $R_\text{Mik}$, which is affected by deviations in the Rabi frequency and detuning, and a family of unitaries: 
\begin{equation}
    \hat{R}(\alpha,\pi/2,\beta)=\hat{R}_z(\beta)\hat{R}_x(\frac{\pi}{2})\hat{R}_z(\alpha),
\label{eq:R_mikado}
\end{equation}
which generalize the simple $\pi/2$ pulse by including arbitrary rotations around $z$-axis.
Each unitary in this family transforms the north pole of the Bloch sphere to a point on the equatorial plane. 
We quantify the above-mentioned error with the infidelity measured as:
\begin{equation}
    \label{eq:J_uni_Mik}
    J_\text{uni}^\text{Mik}=\min_{\alpha, \beta}\left[\frac{2}{3}(1-|\text{Tr}({R}_\text{Mik}^\dagger  \hat{R}(\alpha,\pi/2,\beta))/2|^2)\right],
\end{equation}
which is obtained by generalizing the definition of $J_\text{uni}$ in Eq.~(\ref{eq:J_uni}) to the Mikado family in Eq.~(\ref{eq:R_mikado}).

The infidelity $J_\text{uni}^\text{Mik}$ can be further simplified if we parametrize the Mikado unitary as $\hat{R}_\text{Mik}=\hat{R}_z(\beta')\hat{R}_x(\pi/2+\delta\theta)\hat{R}_z(\alpha')$, which relies on Euler's decomposition.
In this parametrization, the error in the Mikado pulse arises from $\delta \theta$.
Thus, $\hat{R}(\alpha',\pi/2,\beta')$ is the nearest unitary in the Mikado family to $\hat{R}_\text{Mik}$, resulting in:
\begin{equation}
    J_\text{uni}^\text{Mik}=
        \frac{2}{3}\sin^2\left(\frac{\delta\theta}{2}\right)=\frac{\delta\theta^2}{6}+\mathcal{O}(\delta\theta^4).
\label{eq:J_uni_Mik_theta}
\end{equation}
In the optimization of Mikado pulses, we minimize the cost function $J^\text{Mik}$, which is obtained by replacing in   Eq.~(\ref{eq:weighted_sumed_J}) the contribution from $J_\text{uni}$ with its generalized form $J_\text{uni}^\text{Mik}$ in Eq.~(\ref{eq:J_uni_Mik}).

Static deviations of the Hamiltonian parameters arise from spatial inhomogeneities of the laser intensity.
These inhomogeneities contribute to site-dependent deviations of both the Rabi frequency $\Omega$ and detuning $\delta$ (so-called probe shift). 
To make Mikado pulses robust against such laser intensity inhomogeneities, we minimize
\begin{equation}
    \braket{J^\text{Mik}}=\frac{1}{N}\sum_i^N J^\text{Mik}(\delta I_i),
\end{equation}
where $J^\text{Mik}(\delta I)$ represents the Mikado infidelity for a given intensity deviation $\delta I$ from the nominal intensity.
In the numerical optimizations presented in the text, we choose $N{=}11$ relative intensity deviations, uniformly distributed in $[-0.025, 0.025]$.

\section{Strontium-88 setup}
\label{appendix:Sr_setup}

The examples provided in the text are based on ${}^{88}$Sr optical qubits.
In general, the results apply to any other trapped atom or ion with an ultranarrow transition.
The example of ${}^{88}$Sr is special because intensity deviations induce a stronger probe shift \mqvgates compared to other atoms and, thus, testing the recoil-free Mikado gates with this atom highlights their relevance and impact.

The atom is driven at a Rabi frequency $\Omega=2\pi\times \SI{20}{\kilo\hertz}$, assuming that a homogeneous magnetic field of $B=\SI{350}{\gauss}$ is applied to the atom to enable the transition \cite{Taichenachev:2006}.
The probe shift is $\delta\Delta=11.7\,\Omega\,(\delta I/I)$. 
We note that even small relative intensity deviations in the range of few percent induce significant probe shifts on the scale of $\Omega$.
The Mikado pulse has a duration of $T=0.825\pi/\Omega$, which is $65\%$ longer than the zeroth-order quantum speed limit $T_\text{QSL}^0$ for a $\pi/2$ pulse.
The cost function used in the numerical optimizations has weights $w_\text{ent}=100$, $w_\text{mot}=10$, $w_\text{uni}=1$, which is designed to strongly penalize entanglement and motional heating to achieve the recoil free condition.
The unitary error is suppressed by optimization of the $\sigma_z$-rotation angles discussed in the text.
Averaging over the optimization interval  $[-0.025, 0.025]$ of relative intensity deviations, we obtain the average cost functions $\braket{J_\text{uni}^\text{Mik}}=
\num{5.61e-5}$,  $\braket{J_\text{ent}^\text{Mik}}=\num{5.02e-5}$,  $\braket{J_\text{mot}^\text{Mik}}=\num{2.47e-7}$ for the optimal Mikado pulse.
It should be noted that while both  $\braket{J_\text{uni}^\text{Mik}}$ and $\braket{J_\text{ent}^\text{Mik}}$ are of the same order of magnitude, $J_\text{ent}^\text{Mik}$ is nearly independent of the intensity deviation $\delta I$, whereas $J_\text{uni}^\text{Mik}$ displays a strong variation, as shown in Fig.~\figref{fig:4}{c}.
To the purpose of quantum computing, it is key that all qubits (i.e., all sites) perform with comparably low value of the infidelity.

\section{The Composite Pulse Scheme}

We discuss the composite pulse scheme, which uses site-dependent $\sigma_z$ rotations in combination with the Mikado pulses to achieve consistently low unitary infidelities $J_\text{uni}$ for all qubits.

We consider an arbitrary target unitary $\hat{U}_\text{g}$ for the qubit and  parametrize it using Euler's decomposition as follows:
\begin{equation}
\hat{U}_\text{g}=\hat{R}_z(\theta_3^g)\hat{R}_y(\theta_2^g)\hat{R}_z(\theta_1^g),
\label{eqn: Euler_appendix}
\end{equation}
where $\{\theta_1^g,\theta_2^g,\theta_3^g\}$ are the gate angles in this parametrization.
Noting $\hat{R}_y(\theta) = \hat{R}_x(-\pi/2)\hat{R}_z(\theta)\hat{R}_x(\pi/2)$, we rewrite the parametrized target unitary as:
\begin{equation}
    \hat{U}_\text{g}=\hat{R}_z(\theta_3^g)\hat{R}_x\left(-\frac{\pi}{2}\right)\hat{R}_z(\theta_2^g)\hat{R}_x\left(\frac{\pi}{2}\right)\hat{R}_z(\theta_1^g).
\label{eqn: decomposition_appendix}
\end{equation}
Importantly, the $\hat{R}_x$ rotations in this expression can be substituted by unitaries in the Mikado family:
\begin{equation}
    \hat{R}_x(\pm\frac{\pi}{2})=\hat{R}_z(-\beta) \hat{R}\left(\alpha,\pm \frac{\pi}{2},\beta\right)\hat{R}_z(-\alpha).
\label{eqn: mikado_appendix}
\end{equation}
Hence, we obtain the composite pulse scheme $\hat{U}_g$ described in  in the text:
\begin{equation}
    \hat{U}_g=\hat{R}_z(\theta_3)\hat{R}\left(\alpha,-\frac{\pi}{2},\beta\right)\hat{R}_z(\theta_2)\hat{R}\left(\alpha,\frac{\pi}{2},\beta\right)\hat{R}_z(\theta_1),
\end{equation}
where
\begin{equation}
    \theta_1=\theta_1^g-\alpha,\quad \theta_2=\theta_2^g-\alpha-\beta,\quad \theta_3=\theta_3^g-\beta.
\end{equation}

As discussed in Appendix~\ref{appendix:Optimization_Mikado}, the actual Mikado pulse $\hat{R}_\text{Mik}$ displays deviations from the unitaries $\hat{R}(\alpha,\pi/2,\beta)$ in the Mikado family because of deviations in the Hamiltonian parameters;
these deviations are responsible for the unitary infidelity $J_\text{uni}^\text{Mik}$.
The static error in the actual Mikado pulse can be accounted for by the parametrization:
\begin{eqnarray}
    \hat{R}_\text{Mik}[\varphi(t)]&=&\hat{R}(\alpha+\delta\alpha,\frac{\pi}{2}+\delta\theta,\beta+\delta\beta)\\
    \hat{R}_\text{Mik}[\varphi(t)+\pi]&=&\hat{R}(\alpha+\delta\alpha,-\frac{\pi}{2}-\delta\theta,\beta+\delta\beta).
\label{eqn: mikado_err_appendix}  
\end{eqnarray}
where we introduce the deviation angles $\delta\alpha$, $\delta\beta$ and $\delta\theta$.
We note that $\delta \theta$ is the same error angle appearing in  $J_\text{uni}^\text{Mik}$ in Eq.~(\ref{eq:J_uni_Mik_theta}).
To minimize the unitary infidelity introduced by $\delta\alpha$, $\delta\beta$ and $\delta\theta$, the $\hat{\sigma}_z$-rotation angles $\{\theta_1,\theta_2,\theta_3\}$ are replaced by $\{\Tilde{\theta}_1, \Tilde{\theta}_2, \Tilde{\theta}_3\}$ defined as follows:
\begin{widetext}
\begin{eqnarray}
\left.\begin{array}{rcl}
\Tilde{\theta}_1&=&(\theta_1^g-\alpha)-\delta\alpha +s\arctan\left[\tfrac{\sin\left(\tfrac{\theta_2^g}{2}\right)\sin(\delta\theta)}{\sqrt{\cos^2\left(\tfrac{\theta_2^g}{2}\right)-\sin^2(\delta\theta)}}\right]+\frac{\pi}{2}(s-1)\\
\Tilde{\theta}_2&=&2s\arctan\left[\tfrac{\sin\left(\tfrac{\theta_2^g}{2}\right)}{\sqrt{\cos^2\left(\tfrac{\theta_2^g}{2}\right)-\sin^2(\delta\theta)}}\right]-\alpha-\delta\alpha-\beta-\delta\beta\\
\Tilde{\theta}_3&=&(\theta_3^g-\beta)-\delta\beta+s\arctan\left[\tfrac{\sin\left(\tfrac{\theta_2^g}{2}\right)\sin(\delta\theta)}{\sqrt{\cos^2\left(\tfrac{\theta_2^g}{2}\right)-\sin^2(\delta\theta)}}\right]-\frac{\pi}{2}(s-1)
\end{array}
\right\}
&
\hspace{2mm}\text{if}\hspace{2mm}&J_\text{uni}^\text{Mik}<\frac{2}{3}\cos^2\left(\tfrac{\theta_2^g}{2}\right),
\label{eqn: Z-angle_solution_appendix_1}
\\[2mm]
\left.
\begin{array}{rcl}
\Tilde{\theta}_1&=&\theta_1^g-\alpha-\delta\alpha+\frac{\pi}{2}\\
\Tilde{\theta}_2&=&\pi-\alpha-\delta\alpha-\beta-\delta\beta\\
\Tilde{\theta}_3&=&\theta_3^g-\beta-\delta\beta+\frac{\pi}{2}
\end{array}
\right\}
&
\hspace{2mm}\text{if}\hspace{2mm}& J_\text{uni}^\text{Mik}\geq\frac{2}{3}\cos^2\left(\tfrac{\theta_2^g}{2}\right).
\label{eqn: Z-angle_solution_appendix_2}
\end{eqnarray}
\end{widetext}
which can be derived using standard trigonometry. Here, $s=\pm 1$ can be freely chosen. 

For sufficiently small unitary infidelities $J_\text{uni}^\text{Mik}$ of the Mikado pulse, the condition in Eq.~(\ref{eqn: Z-angle_solution_appendix_1}) applies.
When this is the case, the unitary infidelity  $J_\text{uni}$ of the composite pulse gate completely vanishes.
In other words, the error induced by actual Mikado pulse can be fully corrected with $\hat{\sigma}_z$-rotations.
On the other hand, for larger infidelities $J_\text{uni}^\text{Mik}$, Eq.~(\ref{eqn: Z-angle_solution_appendix_2}) applies.
In this case, the error is only partially corrected, and $J_\text{uni}$ can be significantly reduced below  $J_\text{uni}^\text{Mik}$; however, it will not entirely vanish.

We can draw important insight from the previous results.
The condition for vanishing unitary gate fidelity $J_\text{uni}$ can no longer be satisfied when $\cos^2(\theta_2^2/2)$ goes to zero. This is the case for $\theta_g=\pi$.
Thus, we conclude that the unitary infidelity $J_\text{uni}$ of $\pi$ pulses depends more prominently on $J_\text{uni}^\text{Mik}$.
One straightforward way to achieve vanishing values of the $J_\text{uni}$ even in the case of $\pi$ pulses consists in applying twice a composite pulse gate targeting a $\pi/2$ gate instead $\pi$ gate;
the price to pay is a doubling of the execution time.

\section{Mikado vs.\ Mößbauer gates for Composite Pulses}
\label{appendix:infidelity_colormap}
\begin{figure}[t]
	\centering
	\includegraphics[width=\columnwidth]{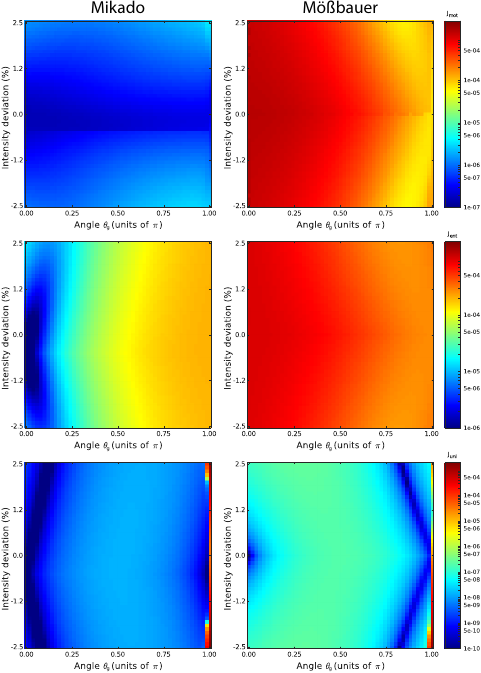}
	\caption{Color map showing the cost functions contributions $J_\text{mot}$ (top), $J_\text{ent}$ (middle), $J_\text{uni}$ (bottom) as a function of the rotation angle $\theta_g$  ($x$-axis) and relative intensity deviation ($y$-axis) for Mikado (left) and Mößbauer (right) composite pulses.
 }
	\label{fig:colormap}
\end{figure}

We analyze the infidelity of the composite pulse scheme for an arbitrary gate $\hat{U}_g$ as defined in Eq.~(\ref{eqn: decomposition_appendix}).
For the global gates, $\hat{R}_x(\pi/2)$ and $\hat{R}_x(-\pi/2)$, which occur in the composite pulse scheme, we consider either an implementation with Mikado gates, $\hat{R}_\text{Mik}[\varphi(t)]$ and  $\hat{R}_\text{Mik}[\varphi(t)+\pi]$, or with Mößbauer gates, defined by a constant phase $\varphi(t)=0$ and  $\varphi(t)=\pi$. 
To make a fair comparison, we improve the fidelity of the Mößbauer gates by rescaling the duration $T$ to account for the dressed Rabi frequency $\tilde\Omega$, which is slightly smaller than the bare Rabi frequency $\Omega$.

Without loss of generality, we restrict the analysis to the family of target gates $\hat{U}_g=\hat{R}_y(\theta_g)$, corresponding to a rotation by an arbitrary angle $\theta_g$ around the fixed $y$-axis.
In fact, according to Eq.~(\ref{eqn: Euler_appendix}), any arbitrary SU(2) gate can be expressed as the product of three rotations around the $z$-, $y$-, and $z$- axes.
Such an arbitrary gate can be realized by adjusting the $\hat{\sigma}_z$-rotation angles $\tilde\theta_1$ and $\tilde\theta_3$ in the composite pulse scheme implementing $\hat{R}_y(\theta_g)$.

For each rotation $\hat{R}_y(\theta)$, we vary the relative intensity deviation $\delta I /I \in [-0.025, 0.025]$, and optimize the three $\sigma_z$-rotation angles for both the Mikado and Mößbauer pulse schemes.
The results of the analysis are presented in Fig.~\ref{fig:colormap}, showing the three cost function terms for both composite pulse schemes in the $(\theta, \delta I/I)$ plane.
The comparison shows that the composite pulses relying on Mikado gates outperform the corresponding pulses derived from Mößbauer gates.
More specifically, the Mikado gates provide an average improvement by nearly three orders of magnitude in $J_\text{mot}$, i.e., in suppressing the photon recoil.
The recoil-free pulses not only suppress motional excitations, but also provide a important improvement in $J_\text{ent}$, which on average is around a factor 15. This improvement becomes even  bigger for small $\theta_g$.
The infidelity associated with deviations from the target unitary, $J_\text{uni}$, also improves for the Mikado pulses, being on average lower by a significant factor 6.
This improvement is attributed to the Mikado pulses, which are computed using optimal control and optimized to be robust against deviations of the Hamiltonian parameters.

\section{Saturation of Entanglement Infidelity in Randomized benchmarking} 
\label{appendix:randomized_benchmarking}
We derive the asymptotic limit of the entanglement infidelity $J_\text{ent}$ in the randomized benchmarking scheme presented in the text.

Each random circuit is defined by a sequence of random unitaries acting on $\mathcal{Q}$,
\begin{equation}
    \hat{U}^c(N)=\hat{U}_g^{(N)}\hat{U}_g^{(N-1)}...\hat{U}_g^{(1)},
\end{equation}
where each $\hat{U}_g^{(n)}$ is sampled from SU(2) according to the Haar measure.
In the simulations, the individual gates $\hat{U}_g^{(n)}$ are implemented based on the composite pulse scheme discussed in the text and illustrated in Fig.~\figref{fig:4}{b}.
Like in Appendix~\ref{appendix:infidelity_colormap}, for the global gates appearing, $\hat{R}_x(\pi/2)$ and $\hat{R}_x(-\pi/2)$, which occur in the composite pulse scheme, we consider either an implementation with Mikado gates, $\hat{R}_\text{Mik}[\varphi(t)]$ and  $\hat{R}_\text{Mik}[\varphi(t)+\pi]$, or with Mößbauer gates, defined by a constant phase $\varphi(t)=0$ and  $\varphi(t)=\pi$.
Note that the implementation of $\hat{U}^{(n)}_\text{Mik}$ and of its counterpart $\hat{U}^{(n)}_\text{Möß}$ are simulated in the larger space $\mathcal{Q}\otimes\mathcal{M}$ space, which includes the motional states, throughout the entire circuit.
Correspondingly, we denote the circuits realized by these pulses as:
\begin{equation}
    \hat{U}^c_\text{Mik}(N)=\hat{U}^{(N)}_\text{Mik}\hat{U}^{(N-1)}_\text{Mik} \ldots \hat{U}^{(1)}_\text{Mik},
\end{equation}
\begin{equation}
    \hat{U}^c_\text{Möß}(N)=\hat{U}^{(N)}_\text{Möß}\hat{U}^{(N-1)}_\text{Möß} \ldots \hat{U}^{(1)}_\text{Möß},
\end{equation}
also acting on $\mathcal{Q}\, \otimes \mathcal{M}$.
Note that the simulation of the quantum circuit keeps track of the motional states throughout the whole quantum circuit and is not reduced to $\mathcal{Q}$ in the intermediate steps.
Hence, the quantum  process tomography is carried out on the entire simulated quantum circuit, using the same procedure developed in the text for the individual gates $\hat{U}_\text{Mik}$.
This approach allows us to quantify the evolution of the three figures of merit, $J_\text{ent}$, $J_\text{uni}$, and $J_\text{mot}$, as a function of the number of gates sequentially applied in the circuit.
The results are presented in Fig.~\ref{fig:5} in the text.

To obtain a quantitative expression for the asymptotic limit of $J_\text{ent}$ in the randomized benchmarking, we use the results derived in Appendix~\ref{appendix:thermal_quantum_channel}.
We make the assumption that we can focus on the first two motional states $\ket{0}$ and $\ket{1}$ and neglect the higher ones.
This assumption is justified by the fact that the initial motional ground state probability is high, $1-p_0 \ll 1$, and that the recoil-free Mikado pulses preserve the probability of occupying the motional ground state.
With this assumption, we can simplify $\hat{U}(T)$ in Eq.~(\ref{eq:unitary_thermal_channel}), which describes the evolution of a recoil-free Mikado pulse in $\mathcal{Q}\otimes\mathcal{M}$, to the following expression:
\begin{equation}
\label{eq:mik_small_temperatures}
\hat{U}_\text{Mik}=\hat{R}_\text{Mik}\otimes\ketm0\hspace{-3pt}\bra{0}+\hat{R}_\text{Mik}\hat{R}_\text{ent}\otimes\ketm1\hspace{-3pt}\bra{1},
\end{equation}
where $\hat{R}_\text{Mik}$ represents an ideal pulse (i.e., an ideal Mikado pulse transforming the north pole to the equator of the Bloch sphere) acting on $\mathcal{Q}$ and $\hat{R}_\text{ent}=\exp{[-i\eta^2\hat{V}_\text{ent}^{(2)}(T)]} $ is derived from Eq.~(\ref{eqn:entanglement_U}) after expressing the entangling unitary as $\hat{U}_\text{ent}=(\hat{R}_\text{ent})^{\hat{a}^\dag\hat{a}}$.
The process described in Eq.~(\ref{eq:mik_small_temperatures}) can be interpreted as follows:
The Mikado pulse is perfectly implemented when the atom is in the motional state $\ketm{0}$, while it suffers from a small unitary error $\hat{R}_\text{ent}$ when the atom is in the state $\ketm{1}$.

The unitary in Eq.~(\ref{eq:mik_small_temperatures}) shows that there are no motion changing terms in the dynamics of recoil-free Mikado pulses.
If we assume, as in the rest of this paper, that the  $\hat{\sigma}_z$-rotations are ideally implemented for all relevant motional states, the motion-preserving dynamics of $\hat{U}_\text{Mik}$ allows us to easily calculate how the atomic state evolves for an increasing number of gates $N$: 
When the atom is in $\ketm{0}$ state, the target circuit $U^c(N)$ is perfectly implemented.
On the other hand, when the atom is in $\ketm{1}$ state, due to the accumulation after each circuit step of an error caused by $U_\text{ent}$ conjugated with random $\hat{\sigma}_z$-rotations, the final qubit state is expected to be fully random with respect to the state evolved under the ideal circuit $U^c(N)$.
Numerical simulations corroborate this assumption.

Thus, the evolution of the entire quantum circuit can be represented as:
\begin{equation}
\hat{U}_\text{Mik}^c(N)=\hat{U}^c(N)\big[I\otimes\ketm0\hspace{-3pt}\bra{0}+\hat{U}_r\otimes\ketm1 \hspace{-3pt} \bra{1}\hspace{-1pt}\big],
\end{equation}
where $\hat{U}_r$ is a random unitary sampled from the Haar measure. By tracing out $\mathcal{M}$, the effect on $\mathcal{Q}$ of the unitary above is the channel $\mathcal{E}(\rho_q)=\hat{U}^c(N)\mathcal{E}_r(\rho_q)\hat{U}^c(N)^\dag$, where
\begin{equation}
    \mathcal{E}_r(\rho_q)=p_0\rho_q+(1-p_0)\hat{U}_r\rho_q \hat{U}_r^\dag.\vspace{2mm}
\end{equation}
To derive the entanglement infidelity, we follow the procedure developed to compute $J_\text{ent}$ in Eq.~(\ref{eq:approx2}) from $\hat{U}_\text{ent}$ in Eq.~(\ref{eqn:entanglement_U}).
For this purpose, it is convenient to parametrize the random unitary as $\hat{U}_r=\exp(-i\theta\hat{\sigma}_\mathbf{n}/2)$, with the rotation axis $\mathbf{n}$ and rotation angle $\theta$ being randomly chosen.
With this parametrization, we thus find that the entanglement infidelity in the asymptotic limit of a large number of gates be expressed as
\begin{equation}
    J_\text{ent}(\theta)=\frac{2}{3}(1-p_0)\sin^2\left(\frac{\theta}{2}\right).
\end{equation}
This expression can be further simplified by considering the distribution of $\theta$, which according to the Haar measure on SU(2) behaves as $p(\theta)=\sin^2(\theta/2)/\pi$.
By averaging $J_\text{ent}$ with this probability distribution, we obtain the following expression,
\begin{equation}
    \langle J_\text{ent}(\theta)\rangle =\frac{1}{2\pi}\int_0^{2\pi}J_\text{ent}(\theta)p(\theta)\,\mathrm{d}\theta = \frac{1}{2}(1-p_0),
\end{equation}
which is the saturation level of $J_\text{ent}$ shown in Fig. \ref{fig:5}. 

\end{document}